\documentclass[useAMS,usenatbib,usegraphicx]{mn2e}

\usepackage{graphics, color} 
\usepackage{hyperref}
\usepackage[all]{hypcap}    
\usepackage{times}
\hypersetup{
    colorlinks,
    citecolor=blue,
    filecolor=blue,
    linkcolor=blue,
    urlcolor=blue
}

\def\enzo{{\sc Enzo}}
\def\moray{{\sc Moray}}
\def\yt{{\sc yt}}

\newcommand\araa{ARA\&A}%
\newcommand\apj{ApJ}%
\newcommand\apjl{ApJ}%
\newcommand\apjs{ApJS}%
\newcommand\aap{A\&A}%
\newcommand\mnras{MNRAS}%
\newcommand\na{New A}%
\newcommand\rmxaa{Rev. Mexicana Astron. Astrofis.}%
\newcommand\ssr{Space~Sci.~Rev.}%
\newcommand\nat{Nature}%
\newcommand\physrep{Phys.~Rep.}%
\def\subsun{\mbox{$_{\odot}$}}
\def\lesssim{\mathrel{\hbox{\rlap{\hbox{%
 \lower4pt\hbox{$\sim$}}}\hbox{$<$}}}}
\def\gtrsim{\mathrel{\hbox{\rlap{\hbox{%
 \lower4pt\hbox{$\sim$}}}\hbox{$>$}}}}

\newcommand\unit[1]{\; \textrm{#1}}

\title[The First Population II Stars]{The First Population II Stars 
  Formed in Externally Enriched Mini-halos}
\author[Britton D. Smith, John H. Wise, Brian W. O'Shea, Michael
  L. Norman, and Sadegh Khochfar]
       {Britton D. Smith$^{1}$\thanks{e-mail:brs@roe.ac.uk},
         John H. Wise$^{2}$, 
         Brian W. O'Shea$^{3}$\thanks{Department of Computational
           Mathematics, Science, and Engineering; Lyman Briggs
           College; and the Joint Institute for Nuclear Astrophysics,
           Michigan State University, East Lansing, MI 48824, USA},
         Michael L. Norman$^{4}$,\newauthor
         and Sadegh Khochfar$^{1}$\\
         $^{1}$Institute for Astronomy, University of Edinburgh, Royal
         Observatory, Edinburgh EH9 3HJ, UK\\
         $^{2}$Center for Relativistic Astrophysics, Georgia Institute
         of Technology, 837 State Street Atlanta, GA 30332, USA\\
         $^{3}$Department of Physics \& Astronomy, Michigan State
         University, East Lansing, MI 48824, USA\\
         $^{4}$Center for Astrophysics and Space Sciences, University
         of California at San Diego, La Jolla, CA 92093, USA}
\begin{document}

\pagerange{\pageref{firstpage}--\pageref{lastpage}} \pubyear{2015}

\maketitle

\label{firstpage}

\begin{abstract}
We present a simulation of the formation of the earliest Population II
stars, starting from cosmological initial conditions and ending when
metals created in the first supernovae are incorporated into a
collapsing gas-cloud.  This occurs after a supernova blast-wave
collides with a nearby mini-halo, inducing further turbulence that efficiently
mixes metals into the dense gas in the center of the halo.  
The gas that first collapses has been enriched to a metallicity of Z
$\sim2\times10^{-5}$ Z$\subsun$.  Due to the extremely low
metallicity, collapse proceeds similarly to metal-free gas until dust
cooling becomes efficient at high densities, causing the cloud to
fragment into a large number of low mass objects.  This external
enrichment mechanism provides a plausible origin for the most
metal-poor stars observed, such as SMSS J031300.36-670839.3, that
appear to have formed out of gas enriched by a single supernova.  
This mechanism operates on shorter timescales than the time for
low-mass mini-halos (M $\le 5\times10^{5}$ M$\subsun$) to recover
their gas after experiencing a supernova.  As 
such, metal-enriched stars will likely form first via this channel if
the conditions are right for it to occur.  We
identify a number of other externally enriched halos that may form
stars in this manner.  These halos have metallicities as high as 0.01
Z$\subsun$, suggesting that some members of the first generation of
metal-enriched stars may be hiding in plain sight in current stellar
surveys.
\end{abstract}


\section{Introduction}
\label{sec:intro}
The physical processes relevant to star formation have conspired to
produce stars of predominantly low mass for most of the history of the
Universe.  The stellar initial mass function (IMF) appears to be
so robust to variations in environment that debate over its
universality primarily concerns whether the statement ``the IMF is
universal'' should include the word \textit{almost}.  Excellent
reviews of this subject have been provided
by \citet{2002Sci...295...82K} and more
recently, \citet{2010ARA&A..48..339B}.  The one notable exception to
the universality of the IMF is the case of metal-free (Population III
or Pop III) stars.  The mid 2000s was a boom for quality reviews of
this subject \citep{2001PhR...349..125B, 2004ARA&A..42...79B,
2005SSRv..116..625C, 2005SSRv..117..445G, 2006jebh.book..239R,
2009Natur.459...49B}.  
In short, neutral metal-free gas cools very inefficiently as it
collapses, resulting in Jeans mass-scale fragments that are of the
order of thousands of M$\subsun$.  Early simulations found that
this resulted in the formation of only one \citep{2002Sci...295...93A,
2002ApJ...564...23B, 2006ApJ...652....6Y, 2007ApJ...654...66O} or
two \citep{2009Sci...325..601T, 2010MNRAS.403...45S} dense cores that
were free to accrete from this massive envelope and grow uncontested.
More recent simulations have found that the disks surrounding the
massive, central core can be unstable to fragmentation, potentially
providing a channel for lower mass Pop III
stars \citep{2011Sci...331.1040C, 2011ApJ...737...75G,
2014ApJ...785...73S}.  However, simulations that follow the late-time 
accretion onto the central object find that they will grow to be 10s to
1000s of M$\subsun$ in final size \citep{2011Sci...334.1250H,
2014ApJ...781...60H, 2015MNRAS.448..568H, Susa14}, making it quite clear that
the Pop III star formation mode is, at a bare minimum, unquestionably
distinct.  Even on the low mass end of the Pop III
IMF, \citet{2015MNRAS.447.3892H} claim that the existing sample size
of low-metallicity star surveys in the Milky Way can already rule out
the existence of metal-free stars below 0.65 M$\subsun$ with 95\%
certainty, barring enrichment via accretion of metals from the
interstellar medium \citep{2014arXiv1411.4189J}.

Following naturally from the idea that the first stars to ever form in
the Universe represent an exception to the universal stellar IMF is
the question of how this necessary transition in star formation modes
took place.  The Universe was forever changed when supernovae from the
first stars created the first metals, whose introduction to
star-forming gas enhanced its ability to cool and fragment.  From this
argument of increased cooling and fragmentation, two ``critical
metallicities'' have been found for yielding a significant departure
from the thermal evolution of collapsing metal-free gas.  At low
densities\footnote{Unless stated otherwise, all densities and number
densities are assumed to be in the proper frame.}, $\sim10^{4-6}$
cm$^{-3}$, fine-structure lines of atomic C
and O allow collapsing gas to continue to cool past the limitations of
metal-free coolants (primarily H$_{2}$) at a metallicity of
$\sim10^{-3.5}$ Z$\subsun$ \citep{2003Natur.425..812B}.  At much
higher densities, $\sim10^{12-14}$ cm$^{-3}$, dust emission can induce
a sharp cooling phase when the metallicity is only $\sim10^{-5.5}$
Z$\subsun$, assuming the dust-to-gas ratio scales with the gas-phase
metallicity \citep{2000ApJ...534..809O, 2005ApJ...626..627O,
2006MNRAS.369.1437S}.  Three-dimensional hydrodynamic simulations have
shown that the cooling phases that occur when these critical
metallicities are reached can indeed trigger fragmentation in their
associated density regimes (\citet{2001MNRAS.328..969B,
2007ApJ...661L...5S, 2009ApJ...691..441S, 2014ApJ...783...75M} for
the atomic and \citet{2008ApJ...672..757C, 2011ApJ...729L...3D, 
2013ApJ...766..103D} for the dust critical metallicity).  However, it
has also been found that the ability of gas to actually fragment
depends sensitively on the initial conditions, no matter what
the metallicity \citep{2007ApJ...660.1332J, 2009ApJ...696.1065J, 
2009ApJ...694.1161J, 2014ApJ...783...75M}.  In particular, without a
small amount of turbulence to create seed perturbations, monolithic
collapse cannot be avoided \citep{2014ApJ...783...75M}.  Thus, it 
seems that further progress requires understanding the true physical
conditions in which the first metal-enriched stars formed.  

This paper is the first in a series that will investigate the range of
scenarios in which star formation producing the near-universal IMF
first occurred.  
Noting the ability of Pop III stars to completely evacuate their host
mini-halos through ionizing radiation and
supernovae \citep{2004ApJ...610...14W, 2008ApJ...682...49W,
2005ApJ...630..675K}, a number 
of recent works have sought to characterize the required conditions
for metal-enriched star formation in terms of (i) incorporation of
metals via hierarchical structure formation into more massive
halos \citep{2007ApJ...670....1G, 2008ApJ...685...40W,
2010ApJ...716..510G, 2012ApJ...745...50W, 2014MNRAS.440L..76S, 
2014MNRAS.438.1669S, 2015arXiv150103212S} or (ii) fallback onto the
original halo after after a recovery
period \citep{2012ApJ...761...56R, 2014MNRAS.444.3288J,
2014arXiv1408.0319R}.  These are both viable scenarios, but because
they typically involve stars forming with metallicities of Z $\gtrsim
10^{-3}$ Z$\subsun$, they are unlikely to be the formation mechanisms
of the most metal-poor stars observed with [Fe/H]\footnote{[Fe/H]
refers to the log$_{10}$ of the iron abundance 
relative to the sun.} $\sim-5.5$ \citep{2013ApJ...762...28N}.  Here,
we report on a third scenario for 
low-metallicity star formation, whereby mini-halos that have yet to form
stars are externally enriched by a Pop III supernova from a
nearby halo.  We note that while this channel has not before been
shown conclusively to give rise Pop II-like stars, it was previously
suggested as a possibility by \citet{2010ApJ...716..510G}, who found
a halo similar to what we describe here but did not have the ability
to follow the collapse and fragmentation.  This is also similar to the
``gravitational enrichment'' scenario proposed
by \citet{2011MNRAS.414.1145M}.  The simulations presented
here are the first of their kind to follow the metal creation,
non-uniform enrichment of the surrounding medium, and eventual
collapse to high density within a 
single, coherent run.  As such, they represent a significant leap
forward in understanding the origins of the first low-mass stars.

The paper is organized as follows.  In Section \ref{sec:methods}, we
describe the simulation code and the setup of the simulations.  We
present the results in Section \ref{sec:results}.  We first describe
the general evolution of the simulation in
Section \ref{sec:evolution}.  We then investigate the process by which
the Pop II star-forming halo is enriched through turbulent mixing with
the supernova blast-wave in Section \ref{sec:enrichment}.  Following
this, we characterize the physical conditions within the star-forming
cloud, including fragmentation, the role of dust, and the velocity
structure of the gas in Section \ref{sec:phys-cond}.  We conclude the
results in Section \ref{sec:andthen} by estimating how often stars
will form via this external enrichment mechanism.  In
Section \ref{sec:discussion}, we discuss the 
implications of these results and some limitations of our work.
Finally, we summarize the results and discuss the ways in
which this work may be extended in Section \ref{sec:conclusion}.

\section{Methods}
\label{sec:methods}
\subsection{Simulation Code}
\label{sec:code}

\subsubsection{The \enzo\ Framework}

We use the \enzo\footnote{\url{http://enzo-project.org/}, changeset
  \texttt{7b5518a63792}} simulation code
\citep{EnzoMethodPaper} for all simulations presented here.
\enzo\ is an open-source, adaptive mesh refinement + N-body
cosmological simulation code that has been heavily used for 
simulating cosmological structure formation over a large range of 
scales.  \enzo\ has been employed by numerous works to study 
high-redshift structure formation, including Pop III star formation
\citep{2002Sci...295...93A, 2005ApJ...628L...5O, 2007ApJ...654...66O,
  2008ApJ...673...14O, 2009Sci...325..601T, 2010ApJ...725L.140T,
  2011ApJ...726...55T, 2012ApJ...745..154T}, low-metallicity star
formation \citep{2007ApJ...661L...5S, 2009ApJ...691..441S,
  2014ApJ...783...75M}, and the first galaxies
\citep{2012MNRAS.427..311W, 2012ApJ...745...50W, 2013ApJ...773...83X, 
  2014MNRAS.442.2560W, 2014ApJ...795..144C}.  All functionality of the
\enzo\ code is 
thoroughly detailed in \citet{EnzoMethodPaper}, but we briefly
describe the most relevant machinery here.  \enzo\ uses a
block-structured adaptive 
mesh refinement (AMR) framework \citep{1989JCoPh..82...64B} to solve
the equations of ideal hydrodynamics in an Eulerian frame, with
multiple hydrodynamic solvers implemented.  For these simulations, we
use the Piecewise 
Parabolic Method of \citet{1984JCoPh..54..115W}.  This is coupled to 
an N-body adaptive particle-mesh gravity solver
\citep{1985ApJS...57..241E, 1988csup.book.....H}.  When ionizing
radiation fields from stellar sources are present (described below in
Sections \ref{sec:chem} and \ref{sec:starff}), we also use the
\moray\ adaptive ray-tracing radiation transport method of
\citet{2011MNRAS.414.3458W}. The AMR framework dynamically creates
(refines) and destroys (de-refines) grid patches at varying levels of
resolution based on multiple criteria.  This enables the study of
astrophysical phenomena with processes occurring over a large range of
spatial and temporal scales.  We list the refinement criteria employed
in this work below in Section \ref{sec:simsetup}.

\subsubsection{Chemistry and Cooling}
\label{sec:chem}

We use a chemistry model that follows the non-equilibrium evolution
of 12 primordial species (H, H$^+$, He, He$^+$, He$^{++}$, e$^-$,
H$_2$, H$_2^+$, H$^-$, D, D$^+$, and HD) following the method
originally described in \citet{1997NewA....2..209A} and
\citet{1997NewA....2..181A}.  This method has been updated to include 
H$_{2}$ formation via three-body reactions
\citep{2002Sci...295...93A}, H$_2$ formation heating, and
collisionally-induced H$_2$ emission \citep{2009Sci...325..601T}.  We
include the effects of ionizing radiation from Pop III stars by
coupling the chemistry network to the \moray\ radiation transport
solver \citep{2011MNRAS.414.3458W}.  We also include optically thin, 
H$_2$ photo-dissociating Lyman-Werner (LW) radiation with an intensity 
decreasing as $r^{-2}$ from the source together with the H$_{2}$
``Sobolev-like'' self-shielding model of \citet{2011MNRAS.418..838W}.

Just as primordial chemistry appears simple enough to allow for the
hope that it can be computed entirely without limiting assumptions in
a large, three-dimensional simulation, the introduction of the first
heavy elements by Pop III supernovae adds a level of complexity to the
Universe that crushes that hope.  We calculate the cooling from
gas-phase metals following the method of \citet{2008MNRAS.385.1443S},
which uses multidimensional tables pre-computed with the
photoionization code,
\textsc{Cloudy}\footnote{\url{http://nublado.org/}} 
\citep{2013RMxAA..49..137F}.  For simplicity, we assume the cooling
from metals corresponds to a solar abundance pattern.  When Pop III
stars are present, we use a four-dimensional cooling table that spans
the parameter space of density, metallicity, temperature, and electron
fraction to account for the increased ionization from the radiation
field.  When metal-enriched gas begins to reach high densities, we
switch to a table that covers only density, metallicity, and
temperature, but was computed up to higher densities than the
four-dimensional table.

In addition to gas-phase chemistry and cooling, we also take into
account the contribution of dust grains, which provide an additional
channel for H$_{2}$ formation and an efficient cooling mechanism at
high densities.  The implementation used here closely follows the work
of \citet{2000ApJ...534..809O} and \citet{2005ApJ...626..627O}, with
modifications motivated by \cite{2011ApJ...729L...3D}, and is detailed 
in \citet{2014ApJ...783...75M}.  The primary assumption to note is
that the gas-to-dust ratio scales with the metallicity and that the
grain size distribution and abundances follow the model of
\citet{1994ApJ...421..615P}, as in \citet{2000ApJ...534..809O}.

\subsubsection{Population III Star Formation and Feedback}
\label{sec:starff}

We model the formation and feedback of Pop III stars with the method
presented in \citet{2012ApJ...745...50W}.  During the simulation, we
insert a Pop III star particle into a grid cell where the following
conditions are met:
\begin{enumerate}
\item The \textit{proper} baryon number density exceeds $10^{7}$~cm$^{-3}$.

\item The gas flow is convergent ($\nabla \cdot
  \textbf{v}_{\mathrm{gas}} < 0$).

\item The molecular hydrogen mass fraction, f$_{\mathrm{H2}}
  \equiv (\rho_{\mathrm{H2}} + \rho_{\mathrm{H2^{+}}}) /
  \rho_{\mathrm{b}}$, exceeds $5\times10^{-4}$, (where
  $\rho_{\mathrm{H2}}$, $\rho_{\mathrm{H2^{+}}}$, and
  $\rho_{\mathrm{b}}$ are the neutral molecular hydrogen,
  singly-ionized molecular hydrogen, and the total baryon densities,
  respectively).

\item The metallicity is less than 10$^{-6}$ Z$\subsun$.  If the
  metallicity is greater than this, we will explicitly follow the
  collapse.
\end{enumerate}

These physical conditions are typical of metal-free molecular clouds
nearing the end of the `loitering' phase observed in simulations of
Pop III star formation \citep{2002Sci...295...93A,
  2002ApJ...564...23B, 2006ApJ...652....6Y, 2007ApJ...654...66O,
  2009Sci...325..601T}.  In the simulations here, we adopt a delta
function at M = 40 M$\subsun$ for the IMF of the Pop III star
particles.  When a star particle is created, the mass of the star is
removed evenly from a spherical region containing twice the particle
mass.  Once formed, the star particle isotropically emits ionizing and
LW radiation following the stellar evolution models of
\citet{2002A&A...382...28S} with no mass loss.  Following this, the
main-sequence lifetime of the 40 M$\subsun$ Pop III star is 3.86 Myr.
Furthermore, we treat the star as a radiation point source, whose
spectrum is discretised into four energy groups with the following
photon energies and luminosities: (i) hydrogen ionizing photons with
$E = 28 \unit{eV}$ and $L_\gamma = 2.47 \times 10^{49} \unit{s}^{-1}$;
(ii) helium singly-ionizing photons with $E = 30 \unit{eV}$ and
$L_\gamma = 1.32 \times 10^{49} \unit{s}^{-1}$; (iii) helium
doubly-ionizing photons with $E = 58 \unit{eV}$ and $L_\gamma = 8.80
\times 10^{46} \unit{s}^{-1}$; (iv) H$_2$ dissociating photons with
$L_\gamma = 2.90 \times 10^{49} \unit{s}^{-1}$
\citep{2002A&A...382...28S}.  The ionizing radiation field is
transported using the \moray\ radiation transport solver
\citep{2011MNRAS.414.3458W}.  As stated previously, we model the
H$_{2}$ photo-dissociating LW flux to be declining simply as r$^{-2}$
with a self-shielding correction.

When the Pop III star's main-sequence lifetime ends, it explodes
as a standard Type II core-collapse supernova with an explosion energy
of $10^{51}$~ergs.  Following the calculations of
\citet{2006NuPhA.777..424N}, the supernova has a metal yield of
11.19~M$\subsun$ and a total ejecta mass of 38.6~M$\subsun$.  We model
the blast-wave by depositing the explosion energy and mass into a
sphere with a radius of 10 proper pc that has been smoothed at its
surface to improve numerical stability \citep{2008ApJ...685...40W}.
We choose this radius as it marks the end of the free-expansion phase
of the supernova.  At the time of instantiation, the blast-wave is
extremely well-resolved and agrees with the Sedov-Taylor solution.

\subsection{Simulation Setup}
\label{sec:simsetup}

We initialize the simulation with a 500 kpc/h comoving box at $z =
180$ using the \textsc{MUSIC} initial conditions generator
\citep{2011MNRAS.415.2101H} with the WMAP 7 best-fit cosmological
parameters, $\Omega_{m} = 0.266$, $\Omega_{\lambda} = 0.732$,
$\Omega_{b} = 0.0449$, H$_{0} = 71.0$~km/s/Mpc, $\sigma_{8} = 0.801$,
and n$_{s} = 0.963$ \citep{2011ApJS..192...18K}, and using a
\citet{1999ApJ...511....5E} transfer function and second-order
Lagrangian perturbation theory.  We first run an exploratory dark
matter-only simulation with 512$^{3}$ particles to $z = 10$.  We then
use the \textsc{ROCKSTAR} halo finder \citep[][interfaced through the
\yt\ analysis toolkit]{2013ApJ...762..109B} and the
\textsc{consistent-trees} merger tree algorithm
\citep{2013ApJ...763...18B} to locate a halo with total mass (at $z =
10$) of $\sim10^7$~M$\subsun$ and the highest number of unique
progenitor halos with M $\ge 10^{6}$ M$\subsun$.  The second of these
criteria was added to maximize the number of Pop III stars formed, and
hence the amount of metals produced, by the time the halo has reached
z = 10.  The halo selected, referred to from here on as the ``target
halo'', has a total mass at $z = 10$ of $1.7\times10^{7}$ M$\subsun$
and 4 unique progenitors with M $\ge 10^{6}$ M$\subsun$ (and 7 with M
$\ge 5\times10^{5}$ M$\subsun$.)

We generate the initial conditions for the primary simulation using
the same configuration and random seed as the exploratory run,
including baryons as well as dark matter, and with two additional
levels of telescoping refinement around the target halo.  The high
resolution region is a rectangular prism minimally containing all of
the dark matter particles that end up within 3 virial radii of the
target halo.  It has an effective resolution of 2048$^{3}$,
corresponding to a comoving spatial resolution of 0.244~kpc/h, and a
baryon (dark matter) mass resolution of 0.259~M$\subsun$
(1.274~M$\subsun$).

We evolve the simulation from $z = 180$ with the physics described in
Section \ref{sec:code}, allowing adaptive mesh refinement to occur only
within the high resolution region.  Grid cells are adaptively refined
by factors of 2 in each dimension when any of the following conditions
exist:
\begin{enumerate}
\item The dark matter mass within a grid cell is greater than 4 times
  the initial mass (i.e., when more than 4 dark matter particles are
  in the same cell).

\item The gas mass within a grid cell is greater than 4 times the
  mean baryon mass per cell on the root grid multiplied by a factor,
  $2^{-0.2L}$, where $L$ is the refinement level.

\item The local Jeans length is resolved by less than 64 cells.
\end{enumerate}
The negative exponential factor used with the baryon mass criterion
ensures that refinement is mildly super-Lagrangian.  Requiring that
the local Jeans length be resolved by at least 64 cells \citep[a
  factor of 16 higher than the criterion put forth
  by][]{1997ApJ...489L.179T}
has been shown to be necessary to guarantee that fragmentation is
purely physical in origin \citep{2011ApJ...731...62F,
2014ApJ...783...75M}.

We place no strict limit on the maximum number of AMR levels in the
simulation.  However, in practice, the simulation typically reaches 15
levels of refinement before forming a Pop III star, corresponding to a
comoving spatial resolution of 0.03 pc/h.  Finally, we stop the
simulation once 
gas with metallicity of at least 10$^{-6}$ Z$\subsun$ reaches a
maximum number density of $\sim10^{13}$ cm$^{-3}$.  This occurs at 27
levels of refinement, for a spatial resolution of approximately 1.5
AU/h comoving.  As discussed in Section \ref{sec:phys-cond}, after the
first Pop III supernova occurs, we start a second simulation from this
point with all physics the same except without dust grains.

We run the simulations on the Blue Waters supercomputer at the
National Center for Supercomputing Applications on 64 nodes of the
machine, using 8 cores/node.  The primary run took roughly 42 days,
with a total computational cost of approximately $64,000$ node-hours.
For all of the analysis and images presented here, we use the
\yt\footnote{\url{http://yt-project.org}} simulation analysis toolkit 
\citep{2011ApJS..192....9T}.

\section{Results}
\label{sec:results}
\subsection{Qualitative Evolution}
\label{sec:evolution}

\begin{figure*}
  \centering
  \includegraphics[width=0.84\textwidth]{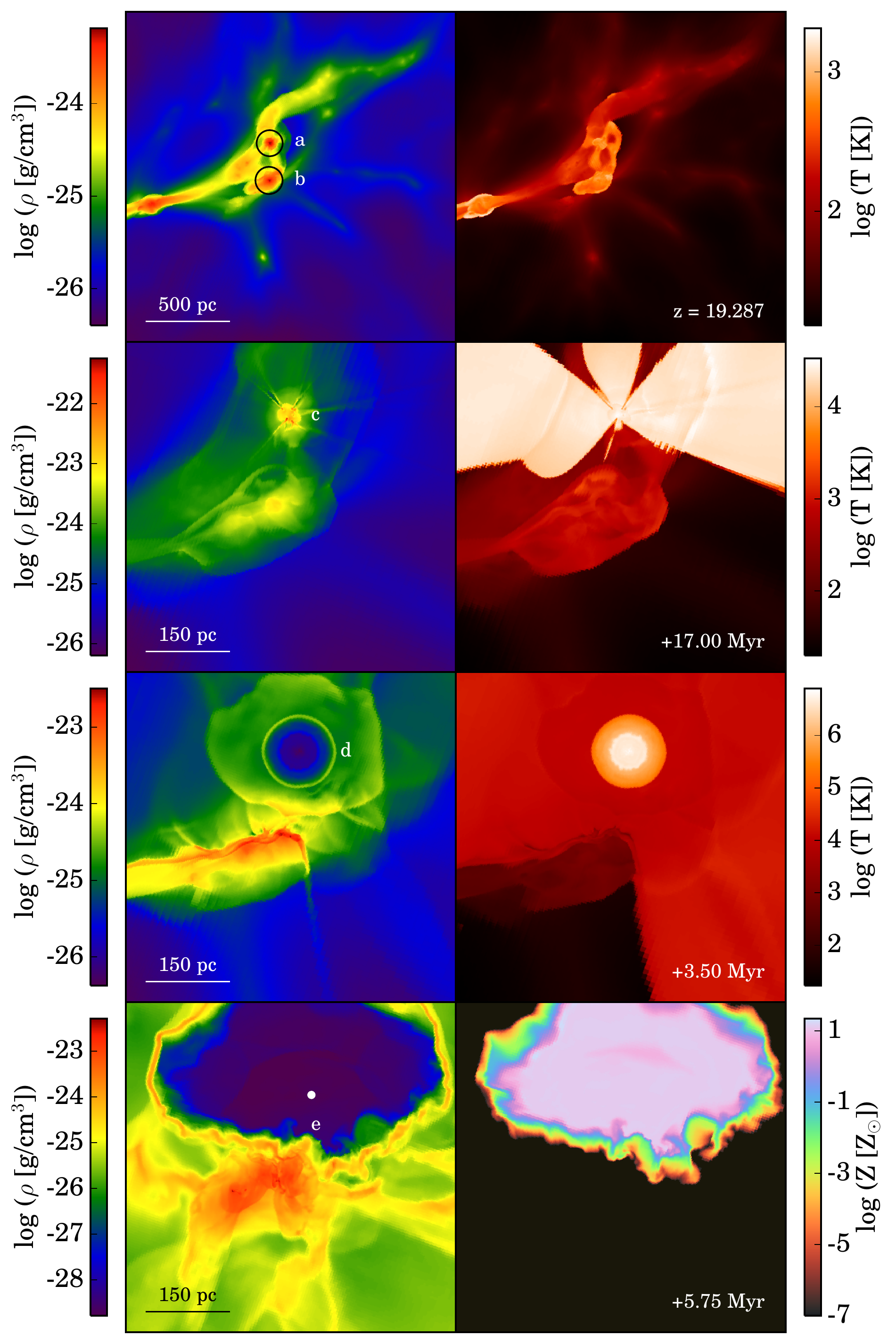}
  \caption{
    Slices illustrating the large-scale evolution of the simulation
    prior to the collapse of metal-enriched gas, oriented such that it
    contains the vector connecting the Pop III star to the rough
    center of the action halo.  Letters indicate the
    following: at $z = 19.287$ a) the Pop III star-forming halo, b)
    the halo in which metal-enriched gas will collapse (i.e., the
    action halo), 
    c) the Pop III star begins shining $\sim 17$ Myr later, d) the Pop
    III star explodes in a core-collapse supernova $\sim 4$ Myr later,
    and e) the supernova blastwave collides with the action halo.  The
    black circles in the top-left panel indicate the virial radii of
    the Pop III (79 pc) and action (82 pc) halos.  The white circle in
    the bottom-left panel shows the location of the evacuated Pop III
    star halo.
  } \label{fig:evolution}
\end{figure*}

\begin{figure*}
  \centering
  \includegraphics[width=0.84\textwidth]{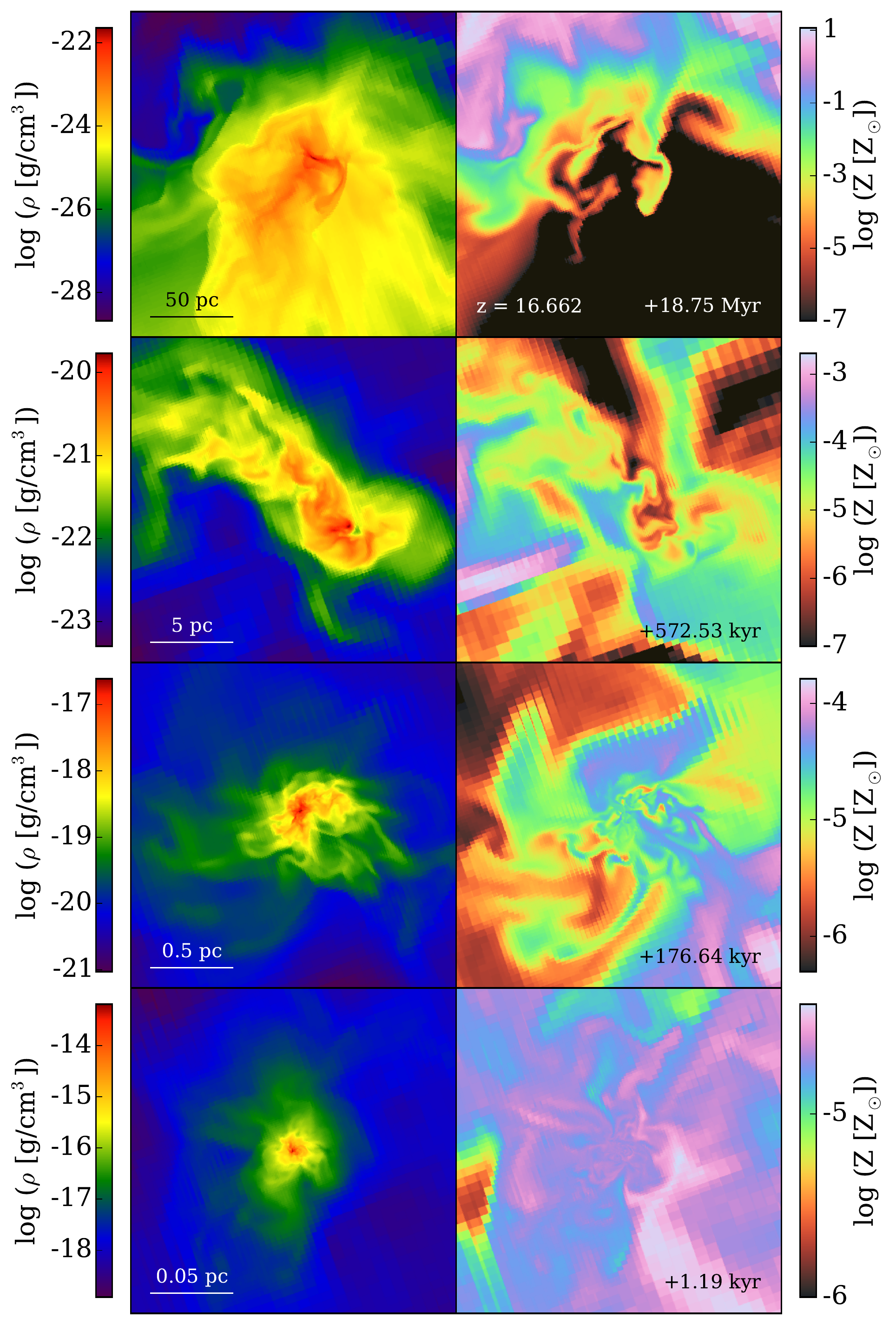}
  \caption{
    Slices showing the turbulent mixing of the metal-enriched
    blastwave with the action halo.  The top panels show the state of
    the simulation $\sim 18.75$ Myr after the bottom panel of Figure
    \ref{fig:evolution}.  Here, as in Figure \ref{fig:evolution},
    the scale bars show distances in the proper frame and the
    slicing plane contains both the Pop III star and
    action halo center.
  } \label{fig:mixing}
\end{figure*}

The two simulations proceed identically until the collapse of
metal-enriched gas first occurs at $z \sim 16.6$, at which time their
behaviors diverge due to the presence/absense of dust grains.  The
qualitative evolution prior to this is as follows.  From here on, we
refer to the halo in which metal-enriched collapse occurs as the
``action halo.''  Within the high resolution region, two Pop III stars
form in separate halos, shine, and explode.  The evolution of the
halos hosting Pop III stars agrees well with earlier studies of this
scenario \citep{2004ApJ...610...14W, 2008ApJ...682...49W,
  2005ApJ...630..675K}.  As the 
halos are only a few hundred thousand M$\subsun$, the Pop III star is
able to completely photo-evaporate them.  The central baryon density
is reduced to roughly 0.1 cm$^{-3}$ before the core-collapse supernova
occurs and totally evacuates the gas from the halo.  As in
\citet{2008ApJ...682...49W}, the supernova shockwave has a density of
just under 1 cm$^{-3}$ and catches up with the expanding shell of the
HII region within a few million years of the explosion.  This
collision enhances the density by a factor of roughly 10 to 30, after
which the shell proceeds to break up into small knots.  These knots
eventually dissolve, but they carry metals out into the intergalactic
medium (IGM), creating
finger-like structures in the metal field, which are clearly visible
in the upper-right panel of Figure \ref{fig:halos}.  The first of the
two Pop III stars, 
forming at $z \sim 23.7$, has no influence on the action halo.  The
second Pop III star forms at $z \sim 18.2$ ($\sim$ 66.6 Myr later) and
is solely responsible for enriching the action halo.  By the time of
gas collapse in the action halo, the supernova blast-waves from each
of the two Pop III-hosting halos (separated by $\sim$ 3.4 kpc) are
still $\sim$ 1 kpc from a collision.  In another version of this
simulation carried beyond this point, we find that the blast-waves
meet at z $\sim$ 14 ($\sim$ 65 Myr in the future).

Figure \ref{fig:evolution} displays the sequence of events immediately
leading up to metal-enriched collapse.  The action halo is roughly 200
pc away from the second Pop III star when it forms.  The Pop III star
photo-ionizes its host halo and causes a slight compression of the gas
in the filament hosting the action halo.  This is visible in the third
row of Figure \ref{fig:evolution}.  Prior to this, the action
halo has a relatively low central baryon fraction, about a few percent
of the cosmic baryon to dark matter ratio.  Over the main sequence
lifetime of the star, the central density of the action halo increases
by a factor of a few, raising the baryon fraction to over 10\% of
$\Omega_{b}/\Omega_{dm}$.  The gas in the action halo remains almost entirely
neutral, with an HII fraction of about 10$^{-4}$ in the central 10 pc,
increasing to 10$^{-2}$ at 50 pc.  Approximately 6 Myr after the
Pop III supernova occurs, the blast-wave encounters the action halo
moving in the opposite direction.  At this point, the action halo has
a virial mass of $\sim3\times10^{5}$ M$\subsun$.  Turbulence produced
by the virialization of the action halo \citep{Wise07, Greif08} and
the passing of the blast-wave allows metals from the supernova to mix
into the core of the action halo.  It also happens that the blast-wave
encounters the action halo just prior to colliding with the HII region
shell and thus meets the halo with an enhanced density.  This process
is illustrated in Figure \ref{fig:mixing}.  Finally, runaway collapse
occurs in the center of the action halo where the metallicity has
reached $\sim2\times10^{-5}$ Z$\subsun$.  While this is happening, the
action halo and the gas-depleted Pop III halo continue on a near
collision course.  By the time collapse occurs, the two halos have
passed close enough that the halo finder considers them to be a single
halo, although distinct cores are still visible.

\subsection{Enrichment of the Action Halo}
\label{sec:enrichment}

\begin{figure}
  \centering
  \includegraphics[width=0.47\textwidth]{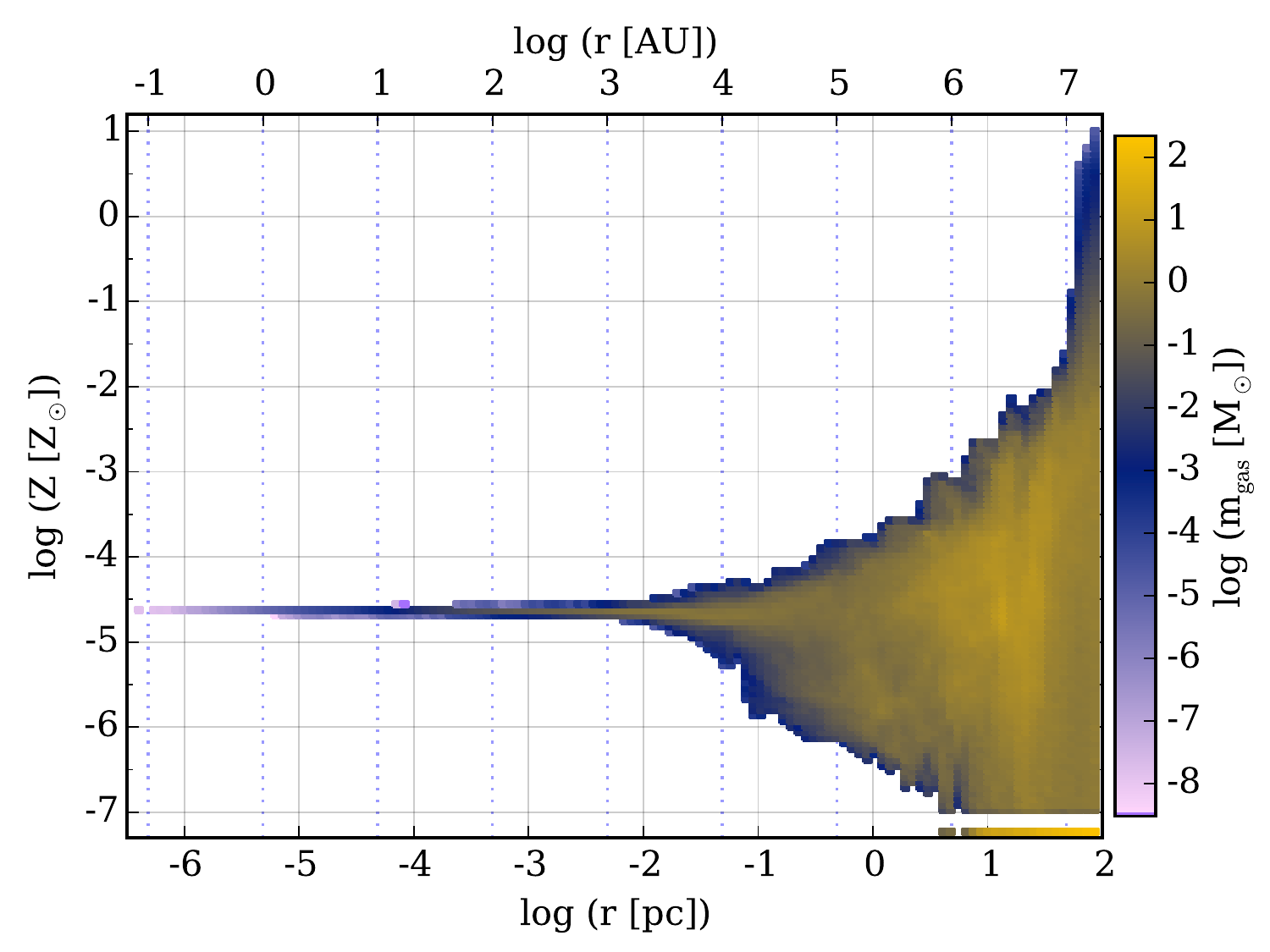}
  \caption{
    Phase plot showing the mass of gas in bins of metallicity and 
    radius from the central core for the simulation with dust.  All
    gas with metallicity Z $<$ 10$^{-7}$ Z$\subsun$ is combined and
    shown in the small strip at the bottom.  This gas is entirely at
    log(r/pc) $>$ 0.5; all gas within that radius has been enriched to
    a higher level.
  } \label{fig:metallicity}
\end{figure}

\begin{figure*}
  \centering
  \includegraphics[width=0.9\textwidth]{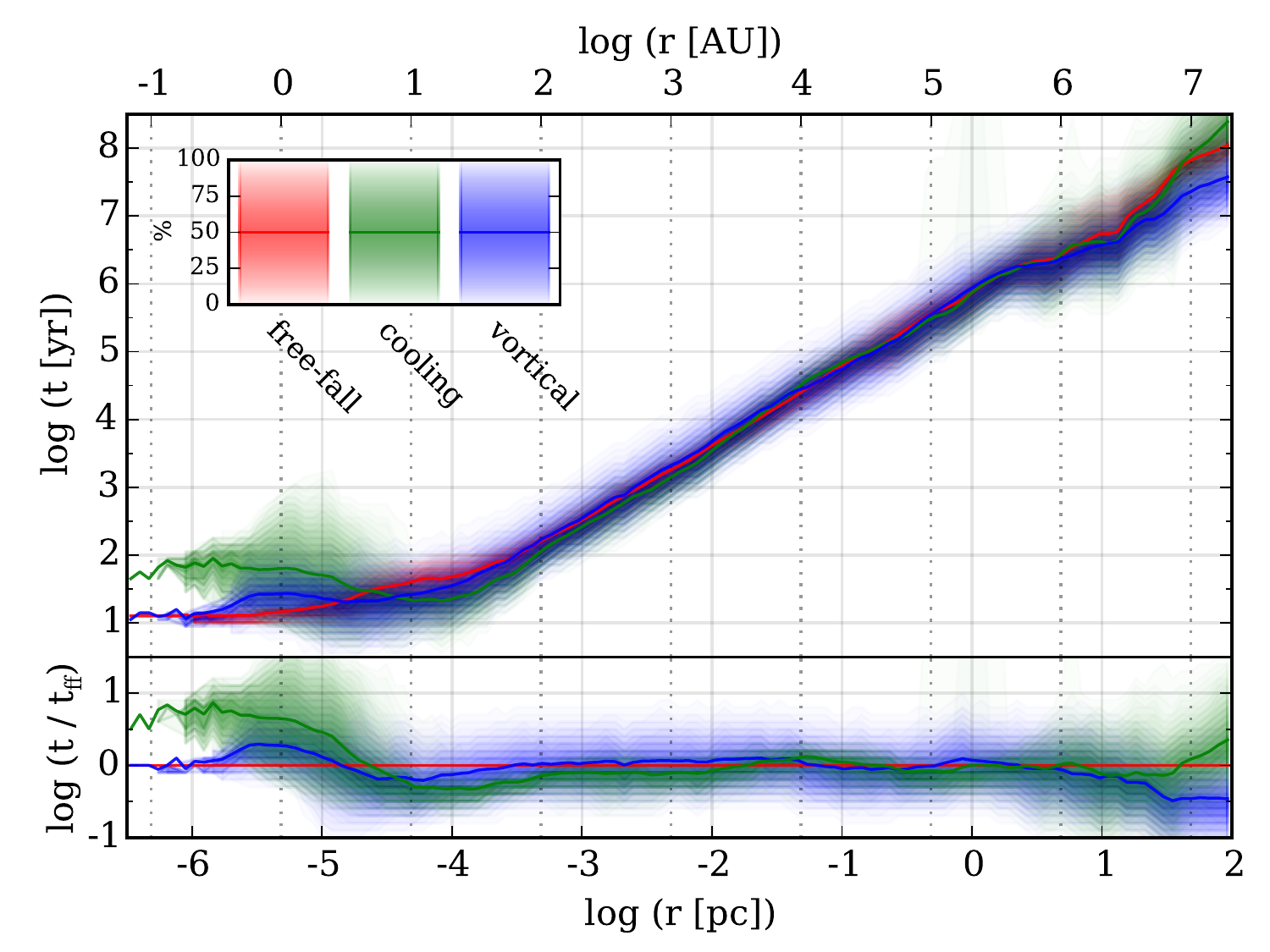}
  \caption{
    Top: radial profiles from the final output of the simulation with
    dust of the mass-weighted mean free-fall time (red), cooling time
    (green), and vortical time (blue).  Bottom: ratio of cooling and 
    vortical times to the free-fall time.
  } \label{fig:timescales}
\end{figure*}

The Pop III supernova blast-wave carries with it 11.19 M$\subsun$ of
metals and impacts the action halo about 6 Myr after the explosion.
Figure \ref{fig:metallicity} shows the metallicity as a function of
radius from the metal-enriched gas cloud at the time of final
collapse.  At the scale of 10--100 pc, gas exists from zero metallicity
up to roughly 10 Z$\subsun$ as spherical shells of this size are
sampling the blast wave itself as well as regions that have yet to be
encountered.  Within a few pc of the dense cloud, no pristine gas
exists and the metallicity is as high as 10$^{-3}$ Z$\subsun$.  Inside
of 0.01 pc, the metallicity is nearly constant at
$\sim$2$\times$10$^{-5}$ Z$\subsun$.  To understand this, we compare
the timescales for collapse and mixing in
Figure \ref{fig:timescales}.  The collapse timescale is the free-fall
time, given by
\begin{equation}
t_{\rm ff} = \sqrt{\frac{3 \pi}{32 G \rho}},
\end{equation}
where $\rho$ is the gas density.  We estimate the mixing timescale as
the vortical time \citep[e.g.,][]{2014arXiv1408.0319R},
\begin{equation}
t_{\rm vort} \simeq \frac{2\pi}{|\nabla \times \textbf{v}|},
\end{equation}
where the factor of 2$\pi$ accounts for a full rotation of an eddy.
The mixing timescale is shorter than the free-fall time outside of a
few pc and, thus, mixing occurs here.  However, at r $\sim$ 1 pc, the
free-fall time is slightly shorter than the vortical time, allowing
the gas to momentarily collapse faster than the rate at which
additional metals can be brought in from the surrounding medium.  We
expect the rate of collapse to be regulated by the free-fall time in
this regime since the cooling time is equivalent to or shorter than
the free-fall time here.

\begin{figure}
  \centering
  \includegraphics[width=0.47\textwidth]{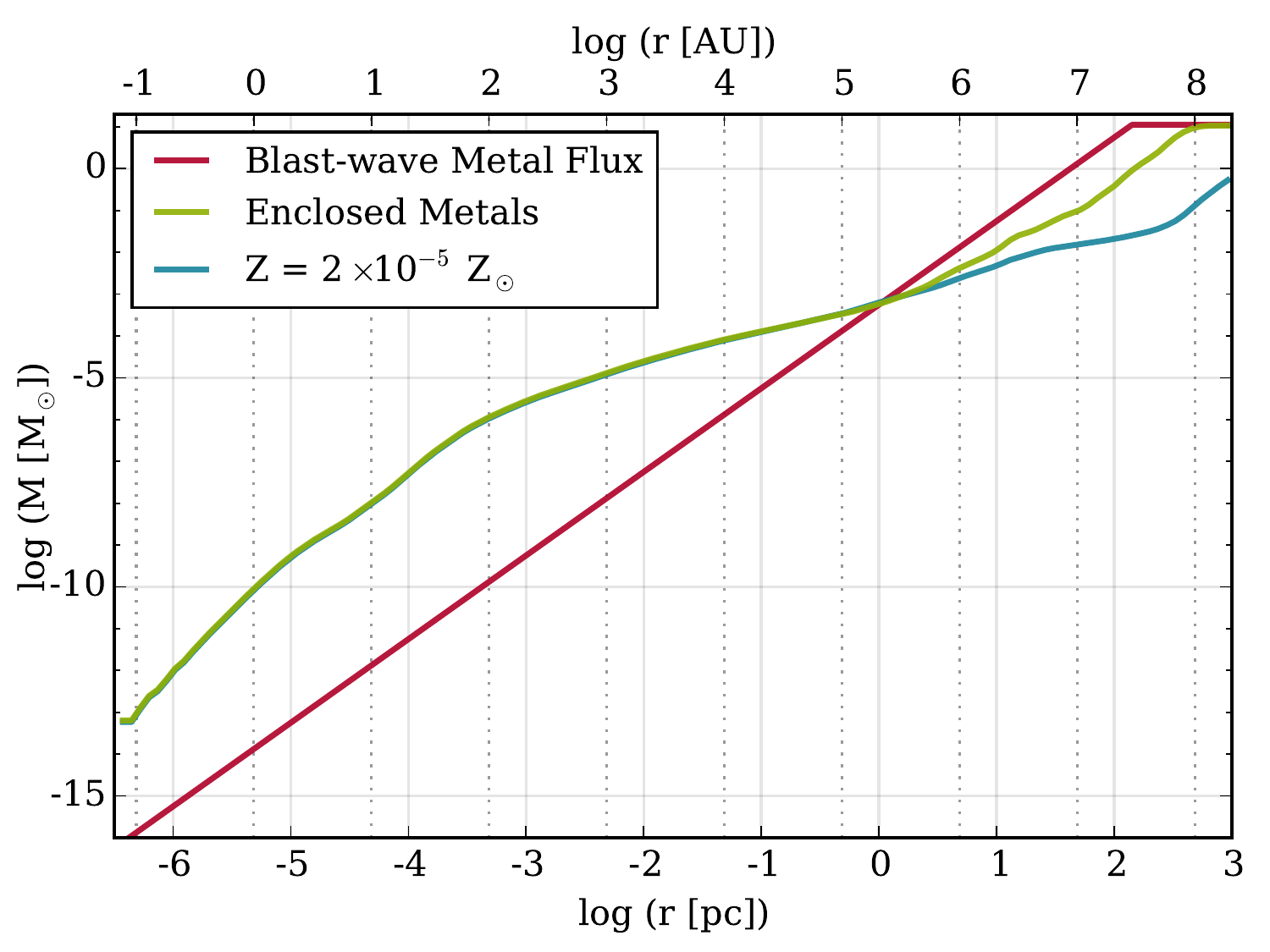}
  \caption{
    The red line shows the maximum metal mass that would have passed
    through a circle of radius, r, in the absence of mixing after the
    blast-wave has reached the action halo.  The green line shows the
    enclosed mass of metals within the cloud and blue shows
    the enclosed metal mass corresponding to a constant metallicity of
    $2\times10^{-5}$ Z$\subsun$.  The flattening of the red line
    indicates the total metal yield of the supernova (11.19
    M$\subsun$).  All of the metals are within 500 pc of the
    center, as shown by the flattening of the green line.
  } \label{fig:metal_mass}
\end{figure}

\begin{figure*}
  \centering
  \includegraphics[width=0.9\textwidth]{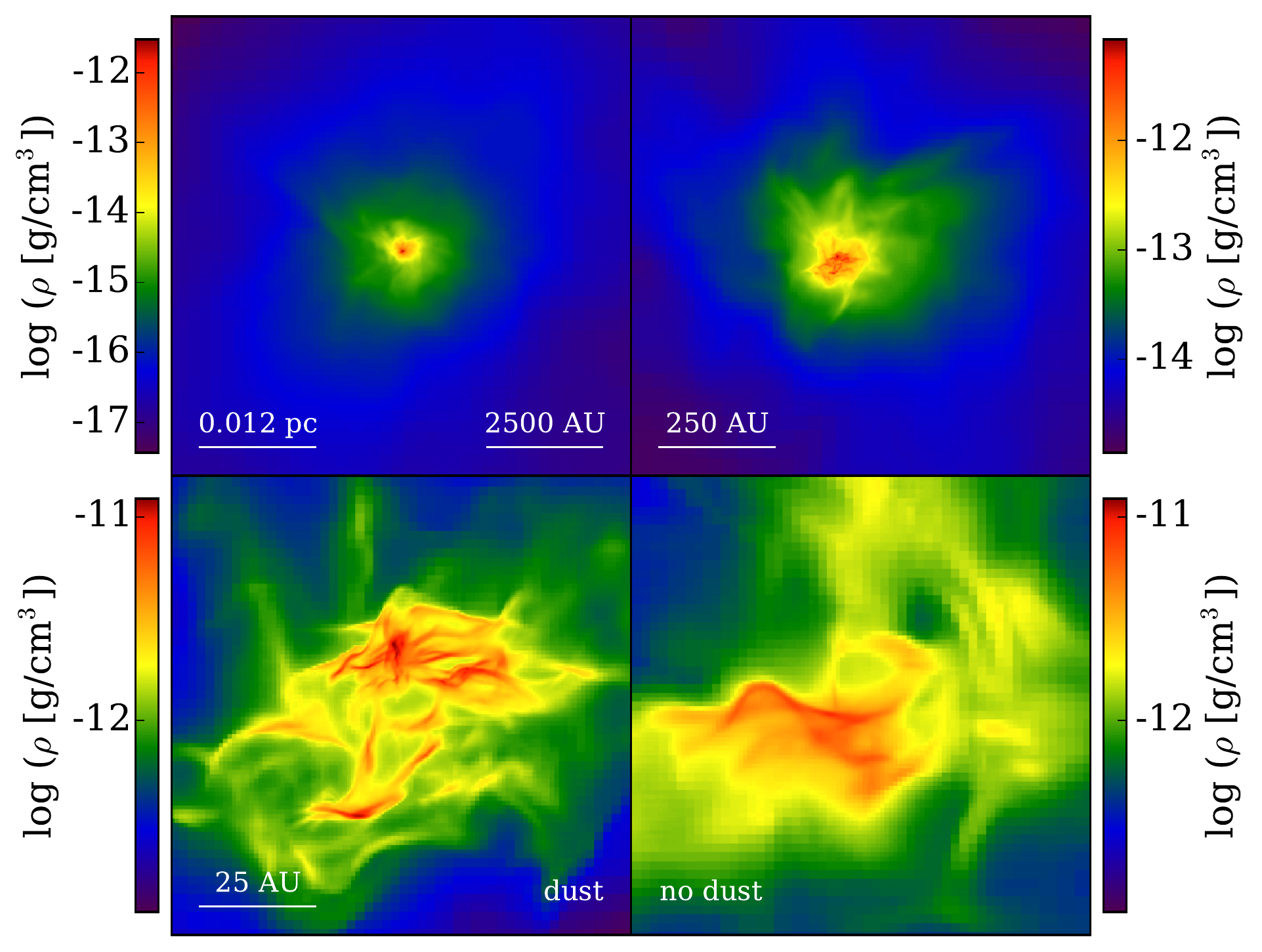}
  \caption{
    Mass-weighted density projections showing the collapse of the
    metal-enriched molecular cloud from the final output of the
    simulations.  The bottom panels show the central, dense core of
    the simulation with (left) and without (right) dust.  Distances
    shown are in the proper frame.
  } \label{fig:collapse}
\end{figure*}

Further evidence of the mixing process can
be seen in Figure \ref{fig:metal_mass}, where we compare the enclosed
mass in metals to the maximum flux of metals carried by the
blast-wave.  Under the assumptions that the blast-wave is spherical and
that all of the metals are carried in the front, the mass in metals
that would have passed through a hole of radius, r, is given by 
\begin{equation}
M_{\rm flux} = M_{\rm SN}\ \frac{r^{2}}{4d^{2}},
\end{equation}
where the metal yield of the supernova, $M_{\rm SN}$, is 11.19 M$\subsun$,
and the proper distance between the halo and the origin of the
blast-wave, d, is $\sim$70 pc.  Where the enclosed metal mass is
greater than the amount carried by the blast-wave, turbulence
has acted to enhance the metal mass over what the blast-wave could have
maximally deposited without mixing.  We also show in
Figure \ref{fig:metal_mass} the mass in metals corresponding to the 
constant metallicity of the inner cloud.  In the region where this
curve diverges from the true enclosed metal mass, mixing is ongoing
and continues to increase the metallicity of the gas beyond the point
at which the metallicity is frozen out by rapid collapse.  We note
here for the sake of novelty that the total mass in metals within 1 pc
of the center of the dense cloud is only about half the mass of
Jupiter.

\citet{2008ApJ...674..644C} study the mixing of metals into
pristine halos impacted by supernova blast-waves using idealized
simulations.  In the configuration most similar to the scenario here, 
(M$_{\rm halo} \sim 3\times10^{5}$ M$\subsun$, v $\sim$ 11 km/s,) they
find that very little mixing occurs with a majority of the dense gas
in the halo remaining at a metallicity less than 3\% of the
blast-wave.  However, there are some key differences between the
conditions we observe in the action halo and the simulations
of \citet{2008ApJ...674..644C}.  The minimum halo mass simulated
in \citet{2008ApJ...674..644C} was 10$^{6}$ M$\subsun$, roughly 3
times more massive than the action halo.  More importantly, the halos
in \citet{2008ApJ...674..644C} were initialized with the gas in
hydrostatic equilibrium with no internal motion.  As
Figures \ref{fig:evolution} and \ref{fig:mixing} show, the action halo
is in a highly perturbed state immediately prior to impact, having
its turbulence first stirred during virialization and then being 
eroded by the Pop III stellar radiation.  This is supported by the
results of \citet{2013ApJ...778...80R}, who perform a similar study to
that of \citet{2008ApJ...674..644C}, but using realistic mini-halos
with M $\sim 10^{6}$ M$\subsun$ extracted from a cosmological
simulation.  Whereas \citet{2008ApJ...674..644C} find that most of the
material in idealized halos of this size remains pristine,
\citet{2013ApJ...778...80R} observe that most of the halo gas is
enriched to $\sim 3$\% of the metallicity of the blast-wave. Finally,
we note 
that while enough metals are mixed into the collapsing gas to
constitute a transition to low-mass star formation in the cloud, the
mixing should not be considered very efficient.  At the 1 pc scale
where free-fall collapse freezes in the metallicity of the dense gas,
Figure \ref{fig:metallicity} shows that the range in metallicity is
nearly 3 orders of magnitude, with metallicities present up to 10
times higher than that of the dense cloud.

\subsection{Physical Conditions of The First Pop II Stars}
\label{sec:phys-cond}

The first Pop II star forming cloud within the simulation has a nearly
constant metallicity of Z $\sim$ 2$\times10^{-5}$ Z$\subsun$.  In
Figure \ref{fig:collapse}, we show density projections progressively
zooming in to the central density peak of the final
output of the simulation.  Not surprisingly, we see no evidence of
fragmentation at densities much lower than the maximum reached in the
simulation, as gas-phase metal cooling has almost no effect at such low
metallicities \citep{2000ApJ...534..809O, 2005ApJ...626..627O,
2009ApJ...691..441S, 2014ApJ...783...75M}.  However, in the central
100 AU, we observe significant fragmentation at densities,
$\rho \gtrsim 10^{-12}$ g/cm$^{3}$, where the cooling from dust
grains becomes efficient \citep{2000ApJ...534..809O, 2005ApJ...626..627O,
2006MNRAS.369.1437S, 2008ApJ...672..757C, 2011ApJ...729L...3D,
2014ApJ...783...75M}.  To confirm that the fragmentation that occurs
is due to the effects of dust, we run a second simulation without dust
starting from the moment following the first Pop III supernova.  In
the lower-right panel of Figure \ref{fig:collapse}, we show the
corresponding central, dense region, which has undergone 
significantly less fragmentation.

\begin{figure}
  \centering
  \includegraphics[width=0.47\textwidth]{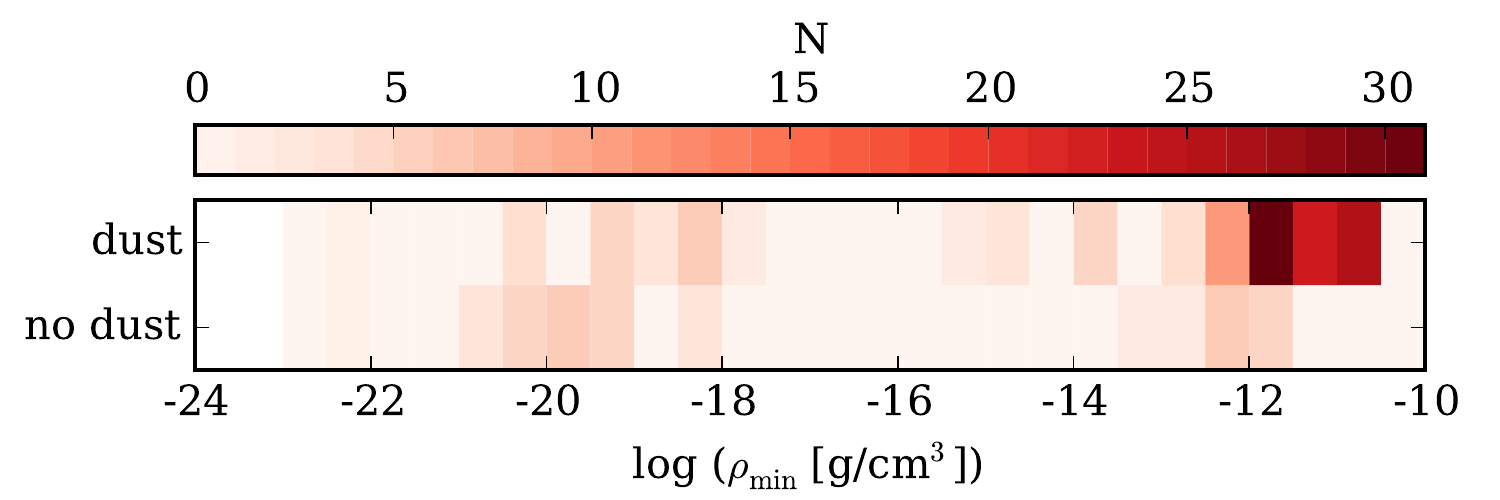}
  \caption{
    The number of bound clumps at a given threshold density (i.e., the
    minimum density of the clump) within the central 
    50 pc of each simulation.
  } \label{fig:clumps}
\end{figure}

To quantify this notion, we run a
clump-finding algorithm to identify bound clumps within 50 pc of the
peak density in both simulations.  Clump
identification \citep{2009ApJ...691..441S, 2011ApJS..192....9T}
proceeds by using a contouring algorithm to calculate the largest
topologically disconnected structures that satisfy the following
criterion:
\begin{equation}
\mathrm{KE} + \mathrm{TE} - \sum_{\rm i} \Lambda_{\rm i}\ t_{\rm dyn,
  i} < \mathrm{PE},
\label{eq:clump}
\end{equation}
where KE, TE, and PE are the total kinetic, thermal, and potential
energies of the clump, and $\Lambda_{\rm i}$ and $t_{\rm dyn, i}$ are the
cooling rate and dynamical time for each member grid
cell \citep{2014ApJ...783...75M}.  As in \citet{2014ApJ...783...75M},
we include the cooling term in Equation \ref{eq:clump} to account for
marginally unbound clumps that are rapidly cooling and, hence,
expected to become gravitationally bound in the near future.  In
Figure \ref{fig:clumps}, we show the number of clumps as a function of
the minimum gas density within the clump.  The minimum density within
the clump effectively serves as the density at which the fragmentation
event forming that clump occurred.  In the simulation with dust, we
identify 126 individual clumps, compared with 38 in the simulation
without dust.  Of the 38 clumps in the simulation without dust, 23 are
at densities less than 10$^{-17}$ g/cm$^{3}$, similar to the
simulation with dust, which has 21 in the same range.  The masses of
these are also similar, ranging from about 0.1 M$\subsun$ up to about
1000 M$\subsun$.  The high density clumps ($\rho_{\rm min} \gtrsim
10^{-13}$ g/cm$^{3}$) in the simulation with dust
show a roughly log-normal distribution in masses, centered at
$\sim10^{-4}$ M$\subsun$ and ranging from about 10$^{-6}$ M$\subsun$
to about 10$^{-2}$ M$\subsun$.  The total gas mass within the central
50 AU is about 0.8 M$\subsun$.  While is is likely that a number of
these may later merge, it is clear that the presence of dust
significantly increases the degree of fragmentation during this phase
of the collapse.

\begin{figure}
  \centering
  \includegraphics[width=0.47\textwidth]{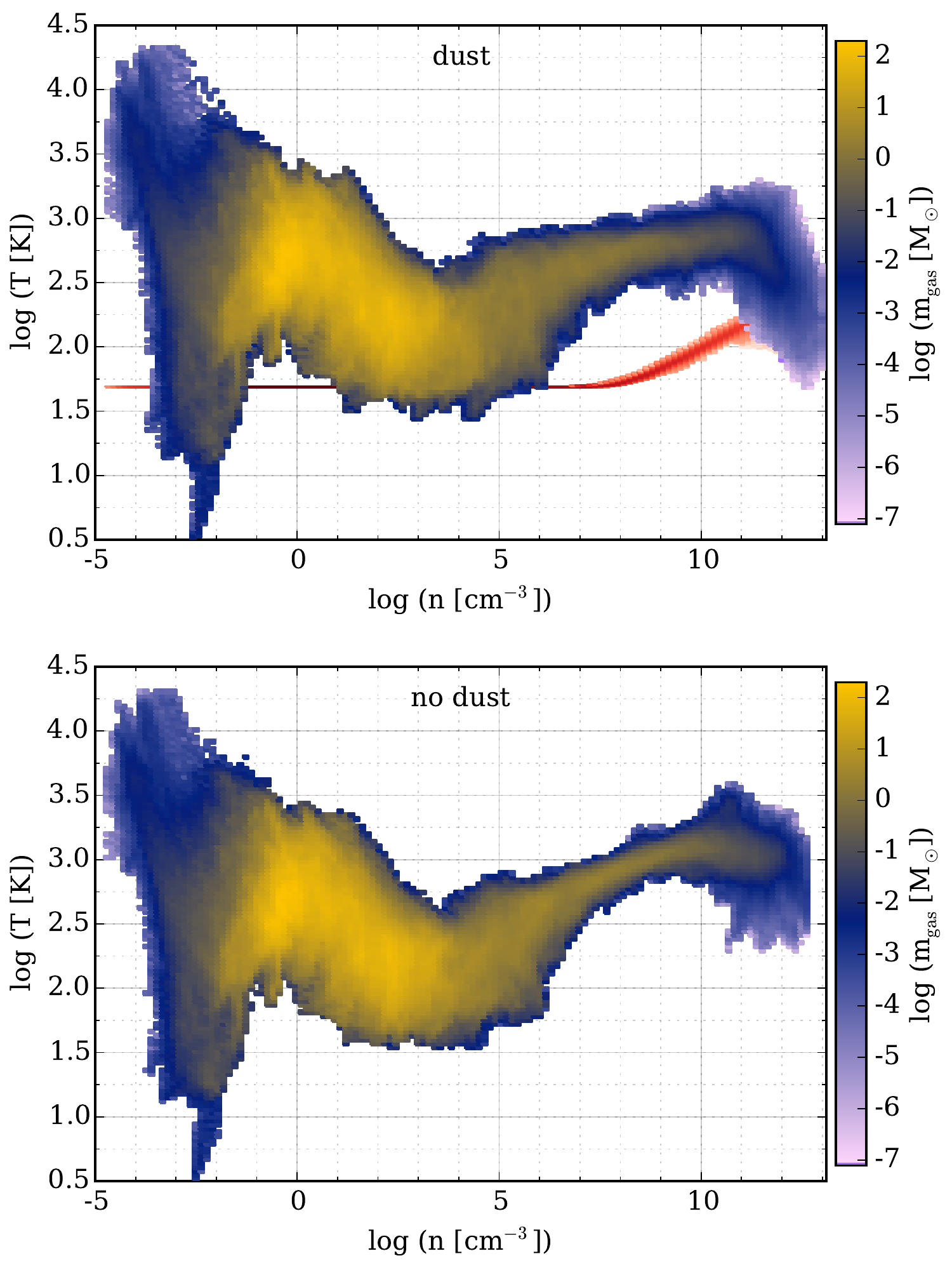}
  \caption{
    Phase plots of the gas mass in bins of number density and
    temperature for the simulation with (top) and without (bottom)
    dust.  In the top panel, the red shows the analogous distribution
    for the dust temperature.
  } \label{fig:phase}
\end{figure}

\begin{figure}
  \centering
  \includegraphics[width=0.47\textwidth]{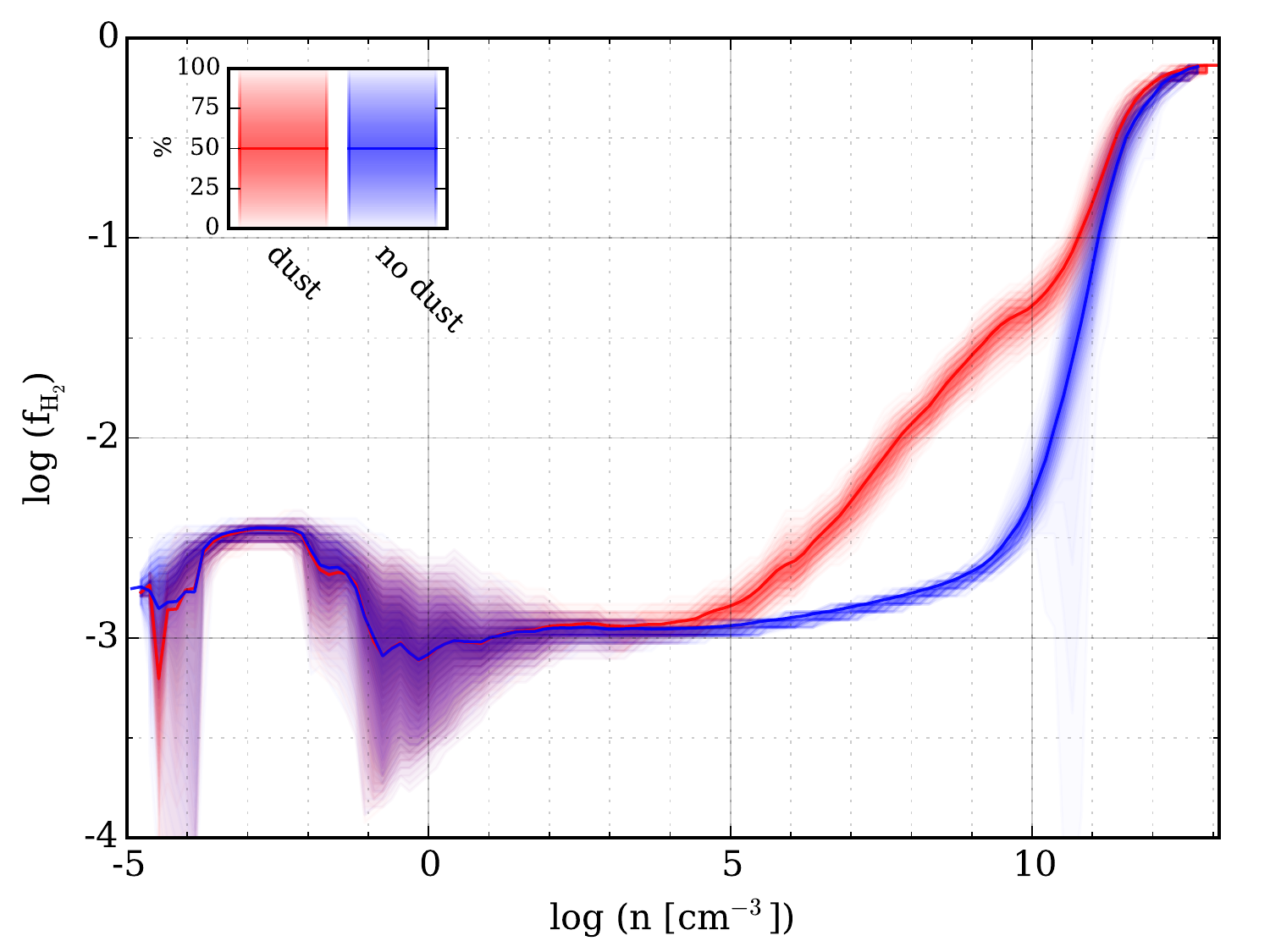}
  \caption{
    Profiles of the mass-weighted average H$_{2}$ fraction as a
    function of number density for the two simulations.
  } \label{fig:fH2}
\end{figure}

The increased fragmenation observed here can be conclusively
attributed to the presence of dust by comparing the thermal state
between between the simulations with and without dust, as we do in
Figure \ref{fig:phase}.  The sharp decrease in temperature seen at
high density, n $\gtrsim$ 10$^{11}$ cm$^{-3}$, occurs when the dust
and gas become thermally coupled and continuum emission from the dust
is able to efficiently cool the gas.  With dust present, gas in the
density range, 10$^{7}$ cm$^{-3}$ $\lesssim$ n $\lesssim$ 10$^{11}$ cm$^{-3}$, is
also cooler by a factor of roughly 2 owing to the increased H$_{2}$
fraction from the availability of the dust-grain formation channel, as
shown in Figure \ref{fig:fH2}.  These results are all in reasonable
agreement with analagous one-zone models \citep{2000ApJ...534..809O,
2005ApJ...626..627O, 2006MNRAS.369.1437S} and three-dimensional
simulations \citep{2008ApJ...672..757C, 2011ApJ...729L...3D,
  2014ApJ...783...75M}.  The 
simulation without dust also shows a decrease in temperature at high
density, albeit much less pronounced.  As Figure \ref{fig:fH2} shows,
this is due to the rapid increase in H$_{2}$ fraction caused by 3-body
reactions becoming efficient.  We note that this appears to happen
somewhat later than in previous works
\citep[e.g.,][]{2005ApJ...626..627O}.  However, it was shown by
\citet{2014ApJ...783...75M} that the precise timing of collapse can 
significantly affect chemical evolution within the gas.

\begin{figure*}
  \centering
  \includegraphics[width=0.9\textwidth]{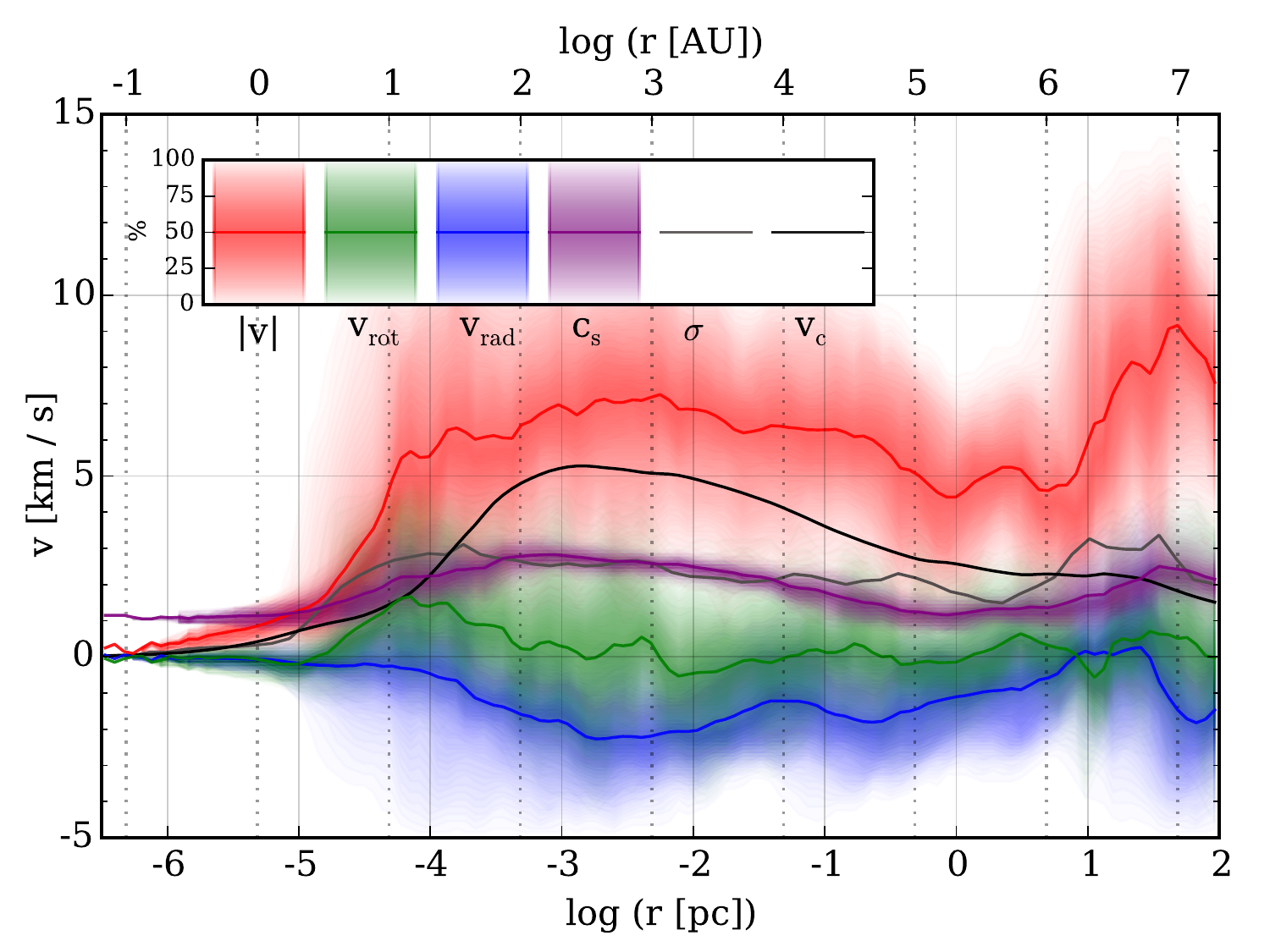}
  \caption{
    Radial profiles of the mass-weighted, mean velocity magnitude
    (red), rotational velocity (green), radial velocity (blue), and
    sound speed (purple).  The variance of the total velocity is shown
    in grey and the circular velocity corresponding to the enclosed
    mass is shown in black.
  } \label{fig:velocities}
\end{figure*}

\begin{figure}
  \centering
  \includegraphics[width=0.47\textwidth]{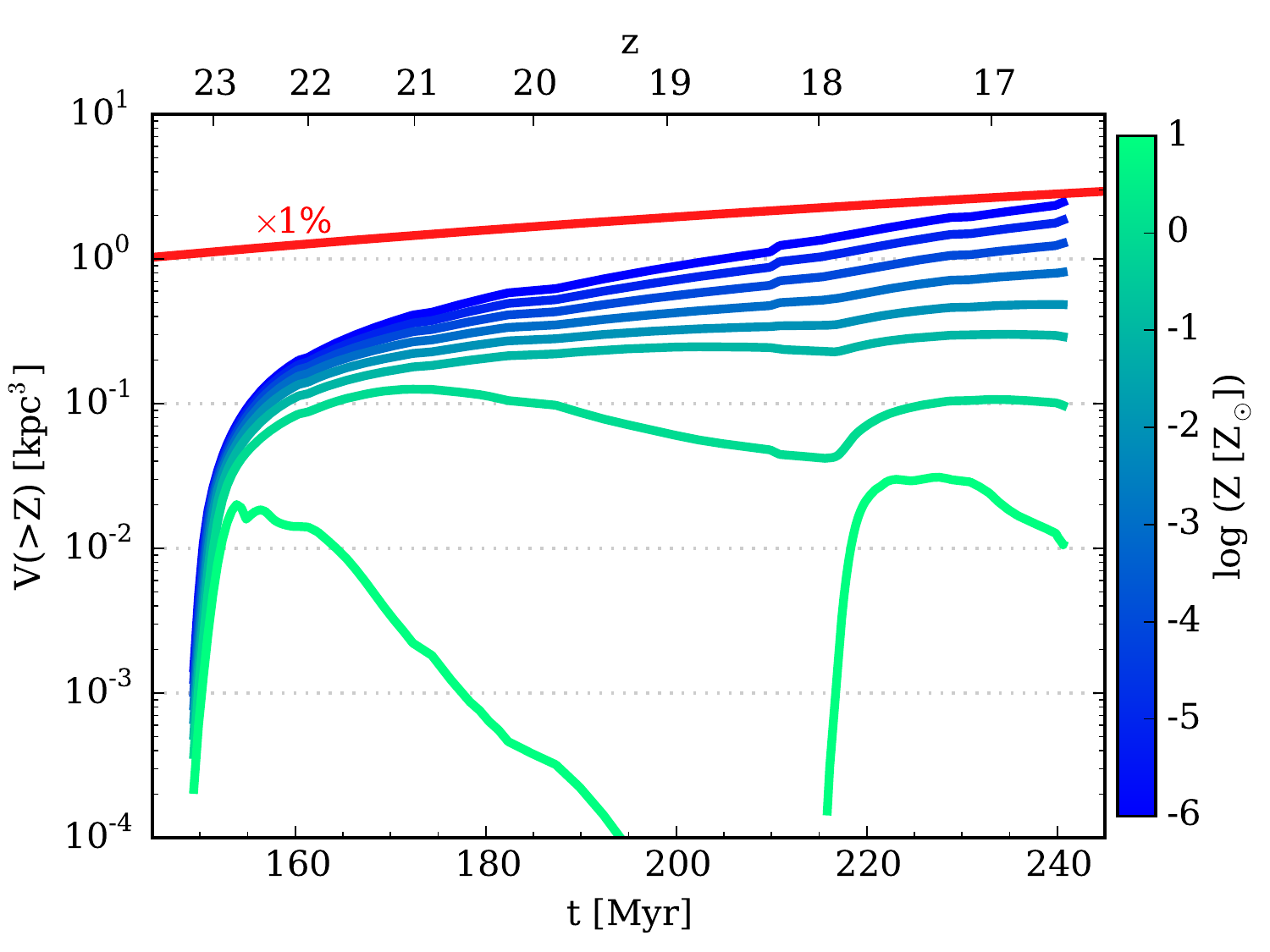}
  \caption{
    The volume within the simulation enriched to
    at least a given metallicity as a function of redshift.  The
    different colors show metallicities ranging from Z = 10$^{-6}$
    Z$\subsun$ (blue) to 10 Z$\subsun$ (green) increasing by factors
    of ten.  The red line shows 1\% of the total volume
    of the high resolution region.
  } \label{fig:metal_volume_profile}
\end{figure}

\begin{figure}
  \centering
  \includegraphics[width=0.47\textwidth]{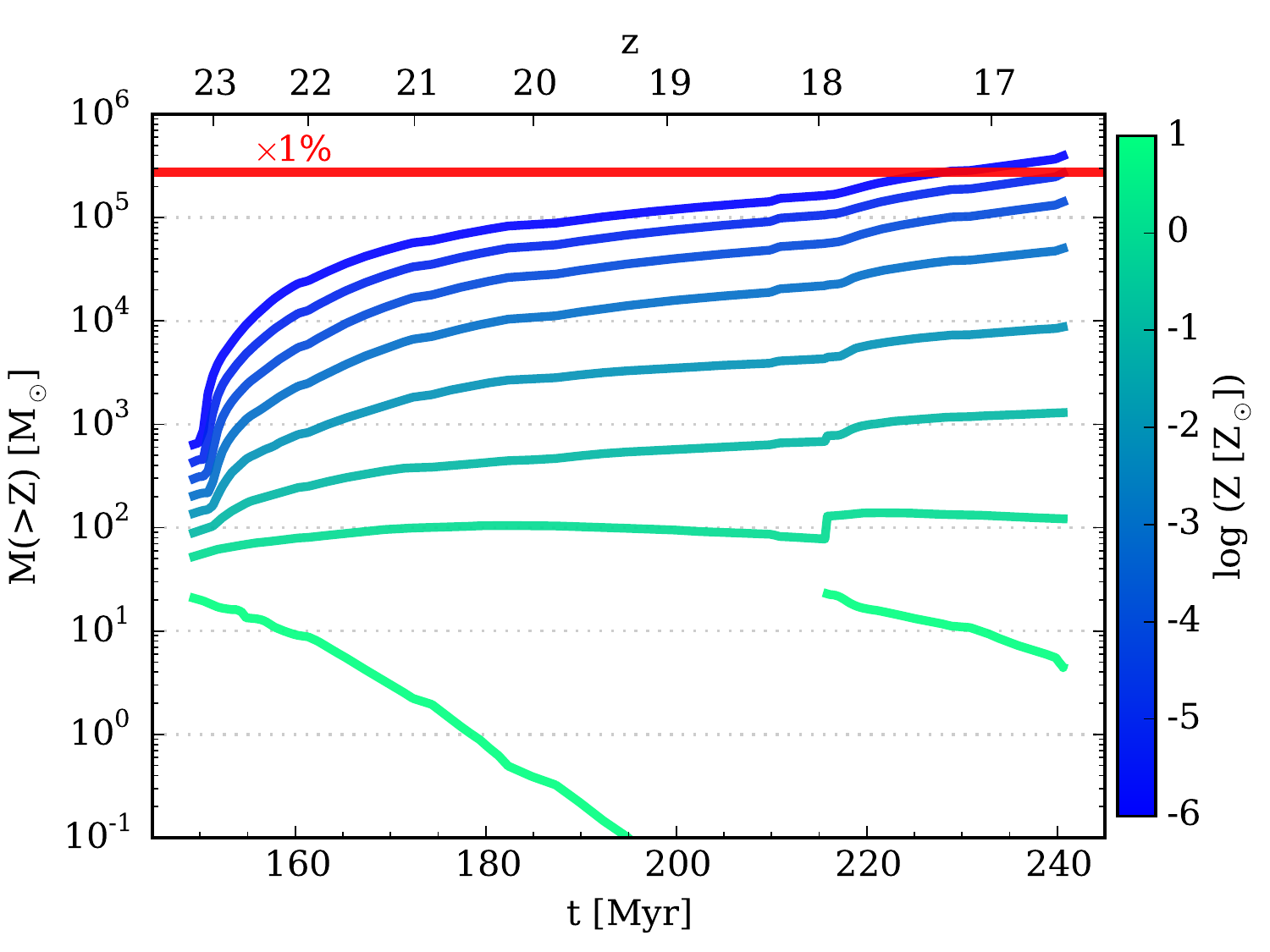}
  \caption{
    The mass of gas enriched to
    at least a given metallicity as a function of redshift.  The
    different colors show metallicities ranging from Z = 10$^{-6}$
    Z$\subsun$ (blue) to 10 Z$\subsun$ (green) increasing by factors
    of ten.  The red line shows 1\% of the total baryon
    mass within the high resolution region.  Note, the refinement
    region is in a large-scale overdensity and has roughly 2.8 times
    more mass than an average region of that size.
  } \label{fig:metal_mass_profile}
\end{figure}

\begin{figure*}
  \centering
  \includegraphics[width=0.9\textwidth]{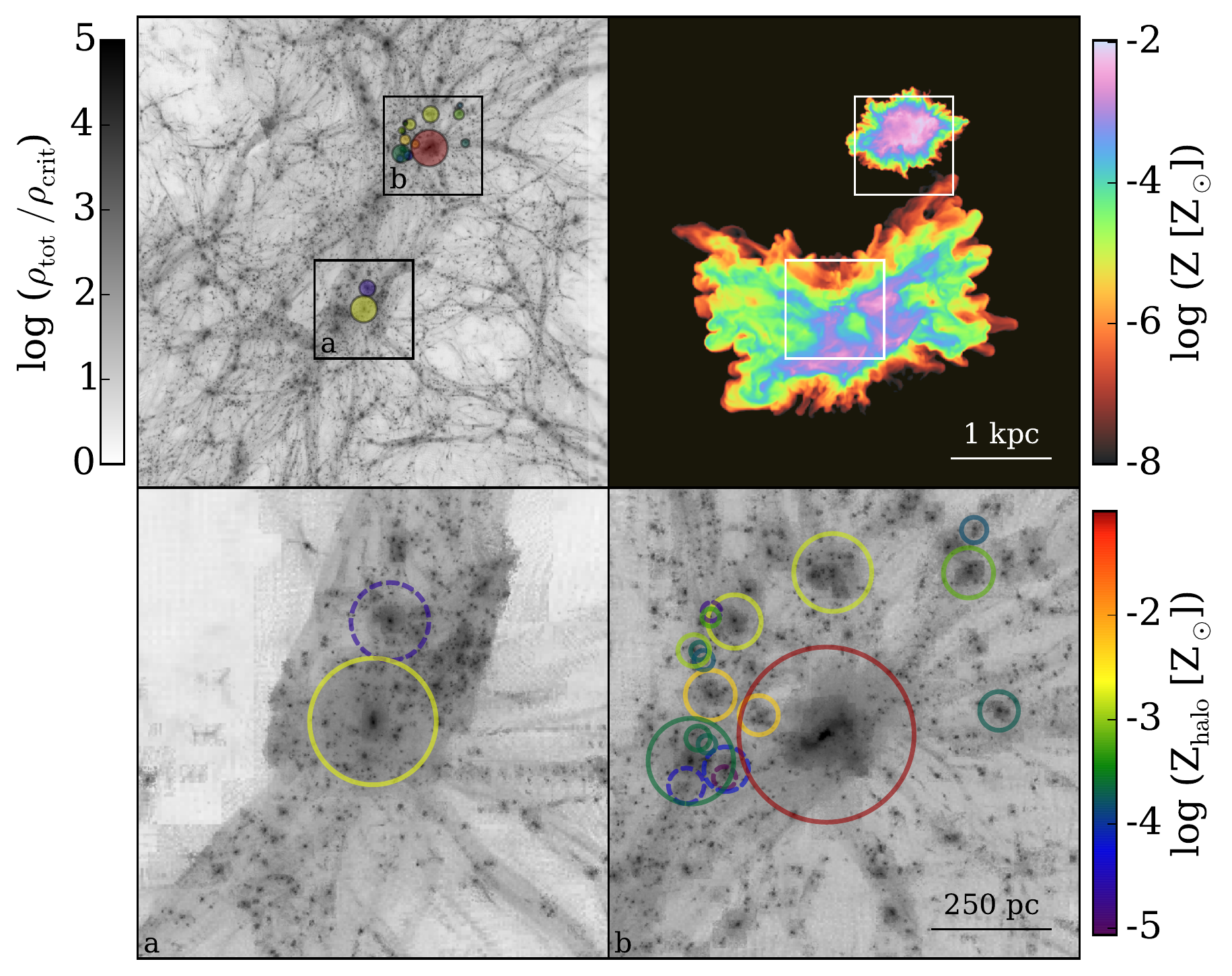}
  \caption{
    Top panels show projections of overdensity (left) and
    metallicity (right) in a region minimally containing all of the
    metals produced in the simulation.  Bottom panels show projections
    of the regions surrounding each of the two supernova events.
    Region a contains the first Pop III star to form and region b
    contains the second as well as the action halo.  The
    locations of halos with metallicities of at least 10$^{-6}$
    Z$\subsun$ in dense gas ($\rho \ge 10^{-25} \unit{g/cm}^{3}$.)  The
    size and color of the circles indicate the virial radii and peak
    dense gas metallicities, respectively.  The dashed lines indicate
    halos with masses less than 10$^{5}$ M$\subsun$ and metallicities
    below 10$^{-4}$ Z$\subsun$ which may be unable to cool.
  } \label{fig:halos}
\end{figure*}

Finally, we briefly examine the velocity structure of the collapsing
gas cloud in Figure \ref{fig:velocities}.  Here, as in Figures
\ref{fig:timescales} and \ref{fig:metal_mass}, we calculate the center
of the cloud by creating a sphere of 100 pc centered on the halo, and
iteratively recalculating the center of mass after decreasing the
radius of the sphere by 10\% until reaching a radius of 1 AU proper.
For the velocity profiles, we subtract off the bulk velocity of the
central 1 AU.  We define the rotational velocity as the component
that is orthogonal to both the radial and angular momentum vectors,
where the angular momentum vector is calculated from the central 1
AU.  The cloud is dominated by turbulent, or at least unordered,
motion as the variance of the velocity magnitude (minus the bulk
velocity of the center) is higher than both the radial and rotational
components at all radii.  At large radii, the velocity variance is
between 1.5 and 2 times the sound speed and closely follows the sound
speed at lower radii, only exceeding it at r $\sim$ 10 AU where the 
cooling leads to rapid cooling.  Though less than the variance at all
radii, there is still a significant infall component that is generally
trans-sonic.  We find that in the simulation without dust where the
gas temperature slightly higher, the infall is similarly raised.  
The gas shows almost no coherent rotation
with an average rotational velocity that is near zero at virtually all 
radii.  This is not simply an effect of where we have chosen to
calculate the angular momentum vector as we find that calculating it
at r = 10 and 100 AU gives qualitatively similar results.

\subsection{The Bigger Picture for Metal-Enriched Star Formation}
\label{sec:andthen}

The gravitational binding energy of mini-halos is well below the
energy of even a standard core-collapse supernova.  As such, it will
be some time before metal-enriched stars are able to form in-situ from
self-enrichment of the progenitor halos that form Pop III stars.
However, just two Pop III stars have enriched a considerable amount of
space and material by z $\sim$ 16.6, when this simulation stops.  In
Figures \ref{fig:metal_volume_profile} and
\ref{fig:metal_mass_profile}, we show the volume and mass enriched to
metallicities high enough to transition to a metal-enriched star
formation mode.  Roughly 1\% of the mass and just under 1\% of the
volume within the high-resolution region of the simulation are
enriched to a metallicity of 10$^{-5}$ Z$\subsun$ by the final output.
Only a small fraction of the metals created in the two supernovae will
be captured by nearby halos and be incorporated into stars by the time
the target halo has assembled at $z = 10$ with a total mass of $\sim$$2
\times 10^7$ M$\subsun$, about 250 Myr from the current time in the
simulation.  If metal-enriched star formation occurs on free-fall
timescales, this 250 Myr time period corresponds to a density of
roughly 10$^{-25}$ g/cm$^{3}$.  Following this assumption, we can
estimate the number of additional halos externally enriched by these
two Pop III supernovae that will be able to form Pop II stars in this
time period by searching for all the halos with metallicities of at
least 10$^{-6}$ Z$\subsun$ in gas with $\rho \ge 10^{-25}$ g/cm$^{3}$.

In Figure \ref{fig:halos}, we show the results of this attempt to
predict the future.  In this exercise, we consider halos with masses
as low as 10$^{3}$ M$\subsun$.  We identify 21 halos in which Pop II
stars could form before the target halo has assembled.  In cases where
the metallicity is less than $\sim10^{-4}$ Z$\subsun$, gas-phase
metal cooling will be insufficient at low densities, requiring a
critical H$_{2}$ fraction in a scenario analogous to the filtering
mass for Pop III star formation \citep{1997ApJ...474....1T,
2003ApJ...592..645Y}.  To account for this, we require halos with Z $<
10^{-4}$ Z$\subsun$ to have a minimum mass of 10$^{5}$ M$\subsun$.  In
other words, halos with M $<$ 10$^{5}$ M$\subsun$ must have
metallicities of at least 10$^{-4}$ Z$\subsun$ such that they are able
to cool via metal lines.  This constraint lowers the number of
candidate halos to 16.  Of these 16 halos, 7 
have masses of 10$^{3}$ M$\subsun <$ M $< 
10^{4}$ M$\subsun$, 6 have masses of 10$^{4}$ M$\subsun <$ M $<
10^{5}$ M$\subsun$, and 3 have masses of 10$^{5}$ M$\subsun <$ M $<
10^{6}$ M$\subsun$.  
Despite having a 66 Myr head start, only 1 of the 16 candidate
halos exists in the region enriched by the first of the two
supernovae.  In fact, this halo is the original host of the Pop III
star.  Out of random chance, the halo hosting the first Pop III star
has significantly fewer neighbors than that of the second Pop III
star.  It has been approximately 96 Myr since the explosion of the
first supernova.  That this halo has not yet recovered is consistent
with \citet{2014MNRAS.444.3288J}, who find a recovery time of 140
Myr for a 3$\times10^{5}$ M$\subsun$ halo experiencing a 10$^{51}$ erg
explosion.  
The lower panels of Figure \ref{fig:halos} show that the halo hosting
the second Pop III star (as well as the action halo) formed in a
significantly denser region with many 
more neighboring halos.  Out of the 22.39 M$\subsun$ of metals
created, about 1.8 M$\subsun$ have been captured by neighboring halos.
It is important to note that as metal-enriched stars form within
region b of Figure \ref{fig:halos}, their feedback may influence other
nearby candidate halos, either inhibiting star formation with
radiation or contributing additional metals via supernovae.  Thus, the
resulting star formation in this region may be slightly different from
the picture presented here, although it is difficult to know for sure
without running the simulation into the future including Pop II star
particles.

What is most striking is the range in metallicities resulting from a
single enrichment event.  A total of 7 halos have metallicities in
their dense gas ($\rho \ge 10^{-25}$ g/cm$^{3}$) of at least 10$^{-3}$
Z$\subsun$, with a single halo 
as high as 0.1 Z$\subsun$.  The halo with the highest metallicity is,
in fact, the combined action/Pop III halo, represented in Figure
\ref{fig:halos} as the large, red circle in the center of region b.
The core of the Pop III halo is just barely visible to the lower-right
of the main core of this structure.  Just after the collision with the
blast-wave but before encountering the Pop III halo, the action halo
experiences a major merger.  The final object, including the not yet
totally merged Pop III halo, has a mass of 9.8$\times10^{5}$
M$\subsun$.  Interestingly, it seems that it will likely form both the
lowest and highest metallicity Pop II stars from the metals created
thus far.  More importantly, the picture presented here shows that 
\textit{the first generation of metal-enriched stars was highly
  heterogeneous in metallicity, encompassing significantly more than
  just the most metal-poor stars.}

\section{Discussion}
\label{sec:discussion}
\subsection{Implications}

The one-zone collapse models with their extremely sophisticated
chemistry networks \citep[][and related works]{2000ApJ...534..809O,
  2005ApJ...626..627O} have paved the way toward understanding the
required conditions for fragmentation from the perspective of the
ability of low metallicity gas to cool.  The simplicity of the
collapse model has enabled the exploration of a large parameter space
of non-solar abundance patterns motivated by Pop III supernova yield
predictions \citep{2006MNRAS.369.1437S} and varying levels of dust
\citep{2012MNRAS.419.1566S}.  Encouragingly, the idealized
three-dimensional simulations \citep{2001MNRAS.328..969B,
2007ApJ...661L...5S, 2008ApJ...672..757C, 2009ApJ...691..441S,
2011ApJ...729L...3D, 2013ApJ...766..103D, 2014ApJ...783...75M}
have confirmed that the thermal 
evolution predicted by these models can, indeed, induce
fragmentation provided there is a small degree of turbulence to seed
density perturbations.  This is particularly interesting given that
the recently discovered most metal-poor star, SMSS J031300.36-670839.3
\citep{2014Natur.506..463K}, has abundance patterns compatible with
the yield of a single Pop III supernova \citep{2014Natur.506..463K,
  2014ApJ...794..100M}.  Additionally, \citet{2014ApJ...794..100M}
have shown that, under a range of dust abundances, a collapsing
gas-cloud with the chemical composition of SMSS J031300.36-670839.3
would have undergone low-mass fragmentation.  What has been missing
from this picture is a means to implant such low metallicities into a
star forming environment.  We have provided such a mechanism in this
work.

\begin{figure}
  \centering
  \includegraphics[width=0.47\textwidth]{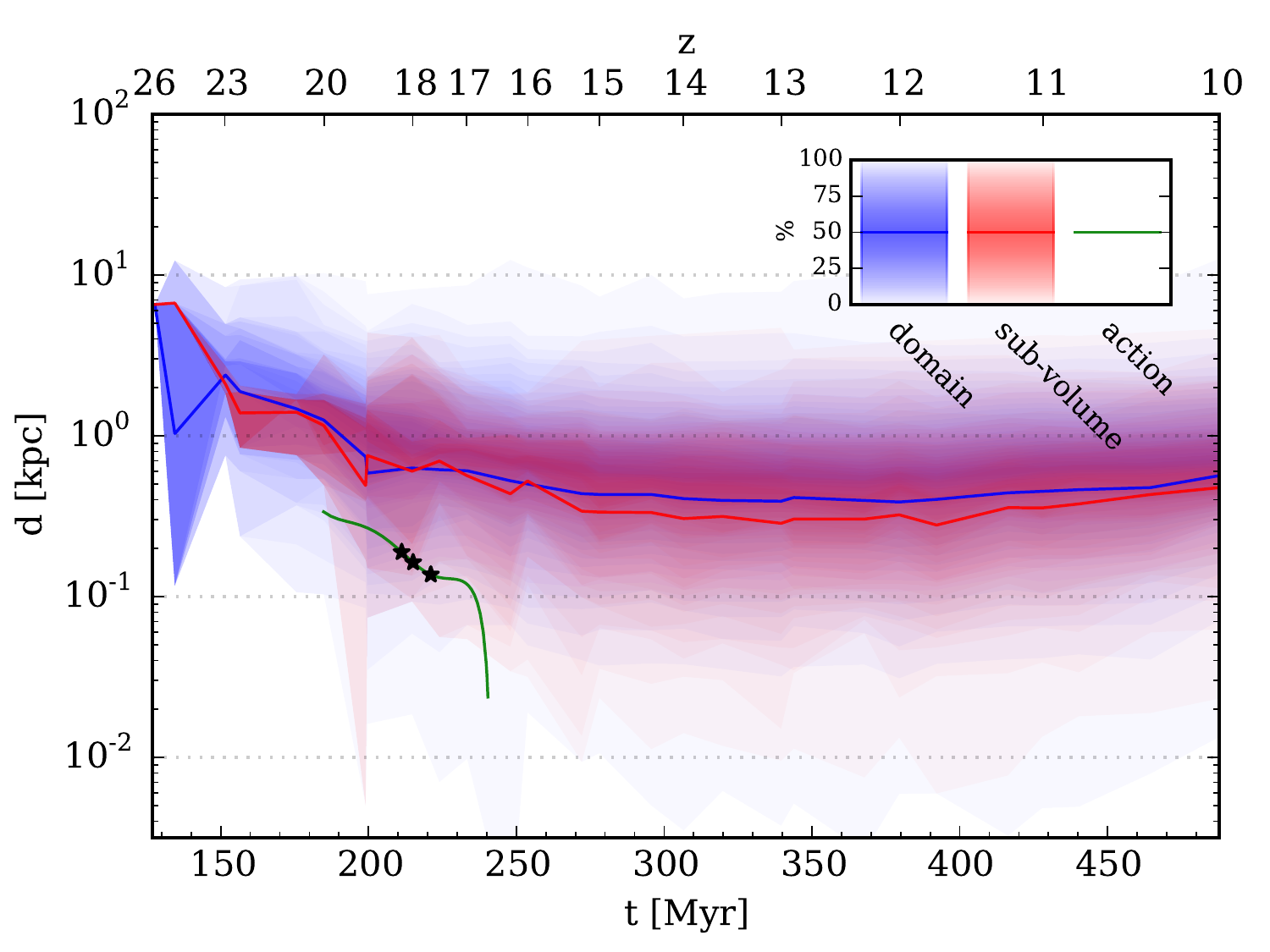}
  \caption{
    The blue line and shaded region shows the average distance between
    halos with mass, M $\ge\ 10^{5}$ M$\subsun$, and the nearest halo
    with mass, 10$^{5}$ M$\subsun \le$ M $\le 5\times10^{5}$
    M$\subsun$ as a function of redshift for the entire simulation
    volume.  The red line and shaded region shows the measurement for
    just the high resolution region.  The green line shows the
    distance between the action halo and the Pop III star-forming
    halo.  The three black stars indicate the moments when the Pop III
    star forms, explodes, and when the supernova blast-wave collides
    with the action halo.
  } \label{fig:neighbors}
\end{figure}

The frequency with which this mechanism operates is uncertain,
although it is also encouraging that \citet{2010ApJ...716..510G}
appear to have identified a similar situation, albeit with a higher
metallicity (Z $\sim2\times10^{-4}$ Z$\subsun$).  We
have, so far, only followed this one occurrence, although we
identify a number of other externally enriched halos that could
potentially form stars with metals from a single supernova.  However,
it is clear that the number of halos externally enriched by a given
Pop III supernova will vary greatly.  For example, the region around
the first Pop III star in our simulation has no viable externally
enriched halo candidates and will see second generation star formation
only after the Pop III halo itself recovers.  By contrast, the region
around the second Pop III star has a number of such candidates.  
To first order, the likelihood of the external enrichment mechanism
depends on the average separation of Pop III star-forming halos from
mini-halos that have yet to undergo star formation.  We provide a
crude estimate of this in Figure \ref{fig:neighbors}.  Using the
exploratory simulation, we identify all halos with masses greater than
10$^{5}$ M$\subsun$ and calculate the distance to the nearest halo
with a mass in the range of 10$^{5}$ M$\subsun \le$ M $\le 5\times10^{5}$
M$\subsun$.  The former group is assumed to have a high probability of
forming a Pop III star, where the latter may be just about to do so.
We allow the mass ranges to overlap noting that the action halo is
actually more massive than the Pop III halo.  During the period from z
= 20 to z = 10, the average separation of these pairs is roughly 500
pc and evolves very little.  The high resolution region of the
simulation shows a similar evolution but with a slightly lower value
of about 300--400 pc owing to the fact that it is located within a
large-scale overdensity.  The distance between the Pop III halo and
the action halo is slightly more than 1 standard deviation lower than
the average pair separation within the full simulation volume.  This
suggests that the external enrichment mechanism is a non-negligible
component of low-metallicity star formation.
\citet{2014MNRAS.444.3288J} find that the recovery time for mini-halos
with M $\lesssim 5\times10^{5}$ M$\subsun$ (similar to the Pop III
star-forming halos in our simulations) is of the order of 100 Myr.
In the simulations presented here, the time between the explosion of
the second supernova and the collapse of metal-enriched gas is
roughly 25 Myr.  The shorter timescale is not surprising, as the
mixing of metals into a nearby halo will occur early on when the
blast-wave is still moving outward quickly.  Thus, if the conditions
are right for the external enrichment mechanism to occur, it should
typically produce metal-enriched stars before they can form from
fallback of metals onto the Pop III halo.

Finally, the candidate halos surrounding the second Pop III star have
a wide range of metallicities, from just over 10$^{-4}$ Z$\subsun$ to
nearly 10$^{-2}$ Z$\subsun$.  This range of metallicities is, itself,
a striking result as it implies that there may be stars that are only
moderately metal-poor by observational standards (with, e.g.,
thousands existing in the SDSS SEGUE sample) yet trace a single
progenitor.  This could 
present a new opportunity to gain access to the Pop III IMF, although
one that will certainly not be without its own challenges.

\subsection{Caveats and Limitations}

We make strong assumptions about the properties of the metals and dust
grains in these simulations.  We assume that gas-phase metals have a
solar abundance pattern and that the dust grains are of a composition,
size distribution, and fraction of total metal mass identical to the
interstellar medium of the Milky Way.  While these assumptions are not
explicitly correct, the thermal evolution we observe is still in
reasonable agreement with studies that use realistic abundance patterns
and dust properties \citep[e.g.][]{2006MNRAS.369.1437S,
  2012MNRAS.419.1566S, 2014ApJ...794..100M}.  We also assume that the
dust grain population is static after the inital grain formation,
although it has been shown recently that grains can continue to grow
during collapse and that their effects can be important even when the
initial depletion fraction is very small \citep{2012ApJ...756L..35N, 
2013ApJ...765L...3C, 2014MNRAS.439.3121C}.  These effects, along with
more realistic abundance patterns, will be important to consider in the
future when these simulations can be carried to the point where an IMF
can be reasonably extrapolated from their final state.  However, given
the current lack of understanding of the physical origin of the
stellar IMF, such an exercise is not warranted at this time.  In
relation to the 
chemistry model employed here, the primary goal of this work is to
show that the high density cooling phase associated with dust can
induce fragmentation in realistic conditions.

We have chosen the formation of a 40 M$\subsun$ Pop III star followed
by a 10$^{51}$ erg core-collapse supernova as the sole outcome of the
collapse of metal-free gas.  Since this work deals with enrichment of
star forming gas by a single supernova, this assumption is as valid as
choosing from a realistic IMF \citep[e.g.][]{2015MNRAS.448..568H}
provided that the mass chosen is plausible within that IMF, which in
this case is true.  However, given the range of alternative Pop III
explosion scenarios, such as energetic hypernova
\citep{2006NuPhA.777..424N} and pair-instability supernova
\citep{2002ApJ...567..532H}, determining the robustness of the
external enrichment mechanism to different explosion energies and
yields has significant merit and is a topic we will explore in a
future study.  For example, \citet{2014ApJ...791..116C} claim that the
lack of observational evidence for Pop III pair-instability supernova
does not necessarily rule out their existence entirely.  Instead, they
argue that the transition to Pop II star formation may simply be
dominated by low energy supernovae whose ejecta fall back quickly
while pair-instability events are masked by the fact that they
evacuate their host halos and have very long recovery times.  However,
if pair-instability supernovae can just as easily give rise to the
external enrichment mechanism presented here, it could present a
problem for this explanation.

Given the nature of our code, which is both grid-based and uses
adaptive mesh refinement, it is possible that the mixing of metals is
artificially enhanced in our second-generation halo.  This is a known
property of grid-based codes 
\citep[e.g.,][]{2008MNRAS.387..427W,2014JCoPh.275..154J}, 
and must be kept in mind when interpreting
results such as those presented in Figure~\ref{fig:metallicity}.  In
this particular simulation, however, there is reason to believe that
numerical mixing is not the origin of the uniform metallicity seen in
the center of our second-generation halo.  This is due to the nature
of the fluid flow in the halo, which is highly turbulent.  Turbulence
results in the efficient mixing of metals over the length scales of
the turbulent cascade on the scale of a few local eddy turnover times,
which are comparable to the local dynamical time in supersonic
turbulence \citep{2005ApJ...634..390B,2010ApJ...721.1765P}.  As can be
seen in Figure~\ref{fig:timescales}, the dynamical, cooling, and
vortical (i.e., turbulence) scales are quite comparable over the inner
parsec of the second-generation halo, suggesting that the gas is
well-mixed.  This thorough mixing may also be observed in nature --
\citet{2012ApJ...759..111K} shows the tentative existence of a
dissolved star cluster with [Fe/H]~$\simeq -2.7$ in the Sextans dSph
galaxy.  This star cluster was found using chemical tagging, and found
a handful of very old stars that are highly clustered in a
multi-dimensional chemical abundance space.  This clumping indicates
chemical homogeneity in the gas cloud out of which these stars formed,
which is estimated to be substantially more massive than the one under
consideration in our calculation, and supports the assertion that gas
should be well-mixed due to turbulent processes.

\section{Summary and Conclusions}
\label{sec:conclusion}
We present the first results from a study designed to characterize the
physical conditions associated with the transition from Pop III to Pop
II star formation.  
We have simulated the formation of the first Pop II stars in the
Universe in a single, coherent simulation beginning from cosmological
initial conditions.  The simulations have a dark matter particle mass
of just under 1.3 M$\subsun$, allowing us to well-resolve the small 
(few$\times10^{5}$ M$\subsun$) mini-halos that host the first stars.
We insert 40 M$\subsun$ Pop III star particles in metal-free molecular
clouds that have evolved past the loitering phase and simulate their
main-sequence lives with radiation hydrodynamics.  Each Pop III star
explodes in a 10$^{51}$ erg core-collapse supernova and we follow the
non-uniform enrichment of the IGM and turbulent mixing that occurs as
the blast-wave washes over neighboring halos.  The simulation ends
when metal-enriched gas is able to cool and collapse into dense,
pre-stellar cores.  The key results from this work are as follows:

\begin{enumerate}

\item The first metal-enriched stars form at z $\sim16.6$ when a single
  Pop III supernova blast-wave washes over a neighboring mini-halo of
  comparable size that has yet to form its own Pop III star.
  Turbulence caused by the virialization of the halo and collision
  with the blast-wave allows metals to reach the interior of
  the halo.  The gas within 0.01 pc of the dense core is uniformly
  mixed to a metallicity of $\sim2\times10^{-5}$ Z$\subsun$.

\item Because the metallicity of the dense core is less than the
  gas-phase critical metallicity (Z$_{cr} \sim 10^{-3.5}$ Z$\subsun$),
  the thermal evolution of the gas as it collapses is similar to the
  metal-free case until dust cooling becomes efficient at n $\sim$
  10$^{12}$ cm$^{-3}$.  At this point, a sharp drop in temperature
  occurs that induces significant fragmentation into many small 
  clumps with masses of roughly 10$^{-4}$ M$\subsun$.  We run a second
  version of this simulation without dust and observe greatly reduced
  fragmentation.

\item The velocity structure of the gas is dominated by turbulent
  motions at all scales, from $\sim$100 pc (the outskirts of the halo)
  down to a few AU.  The turbulent velocity, defined as the variance
  in the total velocity magnitude, roughly follows the local sound.
  At large radii, where efficient mixing occurs, the turbulent
  velocity reaches as high as Mach 2.  The radial velocity also
  follows the local sound speed, but is somewhat smaller.  The gas
  shows virtually no coherent rotation on any scale.

\item By the end of the simulation, two Pop III supernovae have
  exploded in separate regions.  The region in which the first
  supernova explodes has relatively few halos.  As a result, only the
  original host halo of the Pop III star exists as a potential site
  for second generation star formation in this region.  However, 15
  halos exist in the region around the second supernova that have
  metal-enriched gas dense enough to form stars by the time that the
  larger structure encompassing them has assembled at z = 10.  Not counting
  the Pop III host halo, these halos have metallicities ranging from
  slightly over 10$^{-4}$ Z$\subsun$ to nearly 10$^{-2}$ Z$\subsun$.
  Thus, the first Pop II stars may have had a wide range of
  metallicities.

\end{enumerate}

We have identified one pathway, however small, from the exoticism of
the early Universe to the normalcy of the present day.  It is likely 
that multiple scenarios contributed to the transition from Pop III to
Pop II star formation.  While its overall contribution to Pop II star
formation is probably minor, the external enrichment mechanism offers
the best explanation for the origin of the most metal-poor stars
observed and the clearest window into IMF of the first stars.

\section*{Acknowledgments}
We are grateful to the anonymous referee whose comments helped
strengthen the manuscript.  
This research is part of the Blue Waters sustained-petascale computing
project, which is supported by the NSF (award number ACI-1238993) and
the state of Illinois. Blue Waters is a joint effort of the University
of Illinois at Urbana-Champaign and the National Center for
Supercomputing Applications.  The simulations described by this paper
were run using an NSF PRAC allocation (award number
OCI-0832662). B.W.O. was supported in part by the MSU Institute for
Cyber-Enabled Research and the NSF through grant PHY-0941373 and by
NASA through grant NNX12AC98G and Hubble Theory Grant
HST-AR-13261.01-A.  B.W.O was supported in part by the sabbatical
visitor program at the Michigan Institute for Research in Astrophysics
(MIRA) at the University of Michigan in Ann Arbor, and gratefully
acknowledges their hospitality. This research was supported by
National Science Foundation (NSF) grant AST-1109243 to M.L.N.
J.H.W. acknowledges support from NSF grants AST-1211626 and
AST-1333360 and Hubble Theory Grant HST-AR-13895.001-A.  
The analysis presented here made extensive use of new features
developed for version 3.0 of yt.  B.D.S. wishes to thank all the
members of the yt community for their hard work, which helped to
enable this research.  This research
has made use of NASA's Astrophysics Data System Bibliographic
Services.  Computations and associated analysis described in this work
were performed using the publicly available Enzo code and the yt
analysis toolkit, which are the products of collaborative efforts of
many independent scientists from numerous institutions around the
world. Their commitment to open science has helped make this work
possible.

\footnotesize{

}

\label{lastpage}

\end{document}